\documentclass{IEEEtran}
\IEEEoverridecommandlockouts
\usepackage{acronym}
\acrodef{6g}[6G]{six-generation}
\acrodef{mimo}[MIMO]{multi-input multi-output}
\acrodef{siso}[SISO]{single-input single-output} 
\usepackage{amsmath}
\usepackage{graphicx} 
\usepackage{epstopdf}
\usepackage{amssymb}
\usepackage{amsfonts}
\usepackage{booktabs} 
\usepackage{amsthm}
\usepackage{cite}
\usepackage{relsize}
\usepackage{bm,comment}
\usepackage{algorithm}
\usepackage{algorithmic}

\usepackage{subeqnarray}
\usepackage{subfigure}
\usepackage{multicol}
\usepackage{multirow}
\usepackage{diagbox}
\usepackage{slashbox}
\usepackage{stfloats}
\usepackage{float}
\usepackage{color}
\usepackage{cases}
\usepackage{CJK}
\usepackage{bm}

\usepackage[colorlinks,linkcolor=blue,citecolor=red]{hyperref}
\usepackage{lipsum}

\newtheorem{proposition}{\bf{Proposition}}

\newtheorem{lemma}{\bf{Lemma}}

\usepackage{geometry}
\def\BibTeX{{\rm B\kern-.05em{\sc i\kern-.025em b}\kern-.08em
    T\kern-.1667em\lower.7ex\hbox{E}\kern-.125emX}}

\begin{document}
\title{Exposure-Aware Beamforming for mmWave Systems: From EM Theory to Thermal Compliance}

\author{Zihan Zhou, Ang Chen,Yunfei Chen,~\emph{Fellow, IEEE}, Weidong Wang, and Li Chen,~\emph{Senior Member, IEEE}

\thanks{
Zihan Zhou, Ang Chen, Weidong Wang, and Li Chen are with the CAS Key Laboratory of Wireless Optical Communication, University of Science and Technology of China (USTC), Hefei 230027, China (e-mail: zzh31@mail.ustc.edu.cn; chenang1122@mail.ustc.edu.cn; wdwang@ustc.edu.cn; chenli87@ustc.edu.cn).

Yunfei Chen is with the Department of Engineering, University of Durham,
DH1 3LE Durham, U.K. (e-mail: yunfei.chen@durham.ac.uk).
}
}

\maketitle

\begin{abstract}
Electromagnetic (EM) exposure compliance has long been recognized as a crucial aspect of communications terminal designs. However, accurately assessing the impact of EM exposure for proper design strategies remains challenging. 
In this paper, we develop a long-term thermal EM exposure constraint model and propose a novel adaptive exposure-aware beamforming design for an mmWave uplink system. Specifically, we first establish an equivalent channel model based on Maxwell’s radiation equations, which accurately captures the EM physical effects. Then, we derive a closed-form thermal impulse response model from the Pennes bioheat transfer equation (BHTE), characterizing the thermal inertia of tissue. Inspired by this model, we formulate a beamforming optimization problem that translates rigid instantaneous exposure limits into a flexible long-term thermal budget constraint. Furthermore, we develop a low-complexity online beamforming algorithm based on Lyapunov optimization theory, obtaining a closed-form near-optimal solution. Simulation results demonstrate that the proposed algorithm effectively stabilizes tissue temperature near a predefined safety threshold and significantly outperforms the conventional scheme with instantaneous exposure constraints.
\end{abstract}

\section{Introduction}
With the emergence of applications, such as virtual reality, haptic interfaces, and digital twins, communication systems are demanding higher uplink transmission rates \cite{r1}. To meet these requirements, key technologies, such as the use of higher frequency bands, advanced multi-antenna architectures, and beamforming schemes, are expected to be widely deployed \cite{r2}. However, the adoption of these technologies raises concerns over increased human exposure to non-ionizing electromagnetic (EM) radiation and associated thermal hazards \cite{r3}.

To ensure regulatory compliance in EM exposure assessment, two key dosimetric metrics are commonly used.  
One metric is the specific absorption rate (SAR), expressed in watts per kilogram (W/kg), which quantifies the rate at which energy is absorbed per unit mass of tissue \cite{r4}.  
The Federal Communications Commission (FCC) enforces a peak SAR  limitation of 1.6 W/kg for partial body exposure including the head \cite{r5}, 
and the International Commission on Non-Ionizing Radiation Protection (ICNIRP) adopts the whole-body-averaged SAR to 0.4 and 0.08 W/kg for the occupational and public environments \cite{r6}, respectively. The other metric is power density (PD), measured in watts per square meter (W/{m$^2$}), which describes the EM power transmitted through a unit area.  
At high frequencies, EM energy is absorbed superficially within the skin, making incident power density a suitable measure for evaluating surface heating \cite{r7}.  
For instance, IEEE Std C95.1-2019 restricts the incident PD to 20 W/m$^2$ averaged over a 4 cm$^2$ area to prevent excessive tissue heating on the body surface \cite{r8}.  
In light of these regulatory limits, it is imperative to integrate EM exposure constraints into the design of millimeter-wave (mmWave) communication systems.

The modeling of signal-level EM exposure constraints has been widely studied. The early work in \cite{r9} considered an uplink scenario with a two-antenna array and modeled the SAR as a linear combination of transmit power and a cosine term of the inter-antenna phase difference. Subsequent studies generalized this approach by introducing an exposure matrix that encapsulates antenna coupling and spatial directional characteristics, thereby formulating the exposure constraint as a quadratic form of the transmit signal vector \cite{r10,r11,r12}.  Although this mathematical formulation naturally embeds exposure limits into communication system optimization, obtaining the exposure matrix typically relies on extensive full-wave simulation or measured data, resulting in high computational overhead.
To address this issue,  authors in \cite{r13} proposed an efficient EM exposure modeling approach. By incorporating near-field effects and mutual coupling, this method enabled high-precision exposure prediction, providing a concise analytical tool for exposure-aware system design.

Informed by these models, substantial research has been dedicated to maximizing the achievable rate under exposure constraints. For example, a conventional worst-case power back-off scheme was proposed to ensure compliance by scaling down transmit power with a fixed, peak-SAR-derived proportion \cite{r14}. The authors in \cite{r16} formulated the exposure constraint as a quadratic function of the transmit signal and incorporated it directly into the optimization problem.
This allowed the joint design of beamformers under simultaneous transmit power and exposure limits, enabling an adaptive trade-off between regulatory compliance and communication performance.
An EM manifold-based exposure modeling was introduced in \cite{r17}, 
which systematically integrated EM physical principles into the design of exposure-aware beamformers. 
This approach supported finer optimization along the exposure constraint boundary, thereby achieving higher communication performance.  
Further, the authors in \cite{r18} extended this framework to a continuous-time setting and proposed an optimization strategy 
that dynamically allocates the SAR budget over multiple coherence blocks.  By employing dynamic programming and asymptotic water-filling algorithms, 
this method satisfied time-averaged SAR regulations while significantly improving the achievable system rate.

Most of the existing works only focus on instantaneous exposure constraints, i.e., SAR and PD. However, regulatory standards impose long-term exposure constraints at high frequencies. These constraints are specifically designed to prevent thermal hazards arising from prolonged EM exposure \cite{r19}.
For instance, the ICNIRP states that the increase in core body temperature due to EM exposure should be limited to 1$^\circ\text{C}$ over the exposure period to avoid thermal harm \cite{r20}. It also notes that current instantaneous exposure limits are highly conservative relative to this temperature‑based criterion. This is because the human body is an efficient thermoregulatory system, which mitigates significant variations in thermal load induced by EM exposure \cite{r21}.
Therefore, establishing a precise relationship between EM exposure and the resulting thermal effects is of significant importance. The key challenge lies in the fact that predicting tissue thermal effects is highly complex, as it depends on both exposure parameters and intrinsic tissue characteristics, e.g, blood perfusion rate.

To fill in this gap, in this paper, we develop a long-term EM exposure constraint model for an mmWave uplink system and propose a corresponding beamforming design. 
The exposure model is first proposed using the Pennes’ Bioheat Transfer Equation (BHTE) to characterize the thermal response.
The BHTE has been extensively studied in biothermal research, which provides a physically consistent modeling framework by coupling key mechanisms such as EM energy deposition, tissue heat conduction, and perfusion-mediated cooling \cite{r22}. This framework enables us to translate the conventional instantaneous exposure limits into a long-term thermal exposure constraint, defined as a physiological safety temperature threshold during a specified time period. Then, leveraging this novel model, we formulate an adaptive exposure-aware beamforming problem and demonstrate its effectiveness. The main contributions of this paper are summarized as follows:

\begin{itemize}
\item [$\bullet$] \textbf{Long-term EM Exposure Constraint Modeling}: We develop a high-fidelity long-term EM exposure constraint for uplink mmWave systems, where a multi-antenna user equipment (UE) operates near a human head. The constraint is achieved through a two-stage physically consistent modeling process. First, we derive an EM radiation model for the transmit array that incorporates mutual coupling, polarization, and spherical wavefront effects. This model enables the explicit characterization of both near-field exposure metrics and the equivalent communication channel. Second, we establish a dynamic thermal assessment framework based on the BHTE, yielding a closed-form thermal impulse response that captures the thermal inertia of biological tissue. Collectively, this integrated approach translates conventional instantaneous exposure limits into a flexible, long-term thermal exposure constraint.

\item [$\bullet$] \textbf{Adaptive Exposure-aware Beamforming Design}: Building upon the new model, we formulate an adaptive beamforming optimization problem that aims to maximize the long-term average received signal-to-noise ratio (SNR) under the thermal exposure constraint. To solve this problem efficiently, we propose a Lyapunov optimization-based algorithmic framework. In this framework, the long-term thermal constraint is transformed into a virtual queue stability problem, thereby decoupling it into a sequence of per-time-slot subproblems with closed-form solutions. A tunable parameter is also incorporated to explicitly manage the trade-off between communication performance and exposure compliance. Simulation results demonstrate that our proposed scheme outperforms conventional worst-case and per-slot optimal baselines while guaranteeing thermal safety.
\end{itemize}

The rest of this paper is organized as follows. Section \ref{Section2} describes the system model, proposes an EM radiation model, and derives a physically consistent channel model. Section \ref{Section3} further develops an analytical framework for thermal exposure and formulates an exposure-aware beamforming problem. To solve this problem, a Lyapunov-based beamforming algorithm is proposed in Section \ref{Section4}. Numerical results are presented in Section \ref{Section5} to validate the effectiveness of the proposed scheme. Finally, Section \ref{Section6} concludes the paper.

\emph{Notations:} 
In this paper, vectors and matrices are denoted by lowercase and uppercase boldface letters, respectively. We use superscripts $(\cdot)^\mathsf{T}$, $(\cdot)^{-1}$, and $(\cdot)^\mathsf{H}$ to denote the transpose, inverse, and Hermitian transpose, respectively. $\operatorname{diag}(\cdot)$ forms a diagonal matrix from a vector, and $\operatorname{Re}{\cdot}$ extracts the real part of its argument. The symbol $\|\cdot\|$ denotes the $\ell_2$-norm, and $|\cdot|$ represents the modulus operator. $\odot$ stands for the Hadamard (element-wise) product. $\mathbb{C}^{M \times N}$ is the set of all $M \times N$ complex matrices, and $\mathbb{E}(\cdot)$ denotes the mathematical expectation. $\mathbf{I}_N$ denotes the $N \times N$ identity matrix, $\mathbf{0}$ is a zero matrix, and $\mathbf{1}_N$ denotes an all-ones column vector of dimension $N$. $\operatorname{erf}(\cdot)$ denotes the error function. The operator $\nabla \times$ is the curl operator in vector calculus. The imaginary unit is denoted by $j = \sqrt{-1}$, and $\mathcal{C} \mathcal{N}\left(\mu, \sigma^2\right)$ denotes the circularly symmetric complex Gaussian distribution with mean $\mu$ and variance $\sigma^2$.

\section{System Model}\label{Section2}
Consider an uplink mmWave communication system where a UE transmits under the EM exposure constraint. To capture the impact of antenna radiation on both communications and EM exposure, we introduce a geometric model of the system and then derive an EM radiation model that incorporates mutual coupling, polarization, and spherical wavefront. Using this radiation model, a physically consistent channel model is formulated.

\subsection{Geometric Model}
\begin{figure}[t]
\centering
\includegraphics[width=0.47\textwidth]{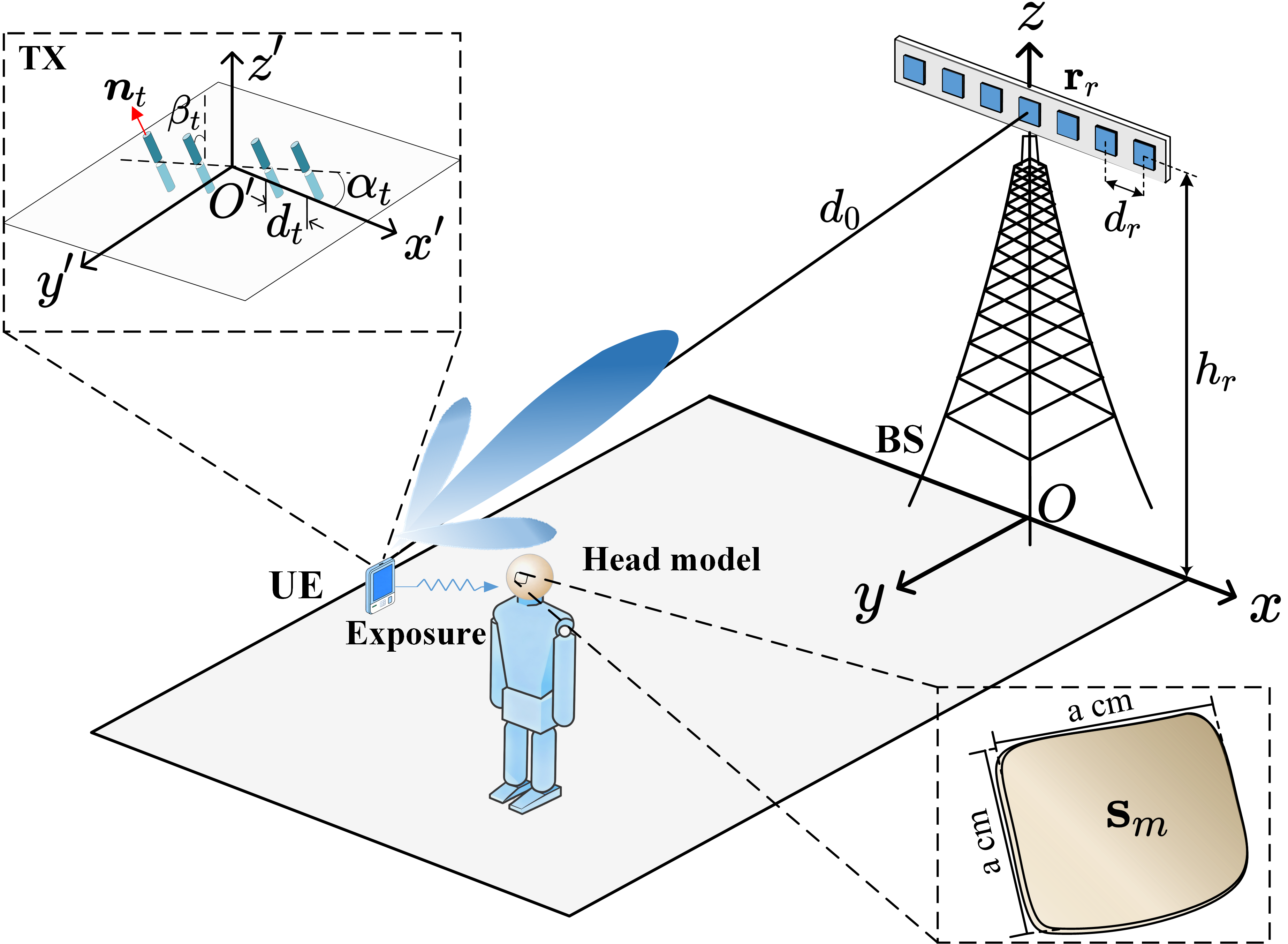}
\caption{Illustration of the mmWave uplink system. A spherical head model of the user lies close to the UE array.}
\label{f1}
\end{figure}

As illustrated in Fig. \ref{f1}, the base station (BS) is equipped with a uniform linear array (ULA) of $N_r$ receive antennas, while the UE employs a ULA of $N_t$ transmit antennas. 
The antenna spacings at the BS and UE are denoted by $d_{r}$ and $d_t$, respectively.
We establish a global Cartesian coordinate system $O$-$xyz$ with unit basis vectors $\hat{\boldsymbol{x}}$, $\hat{\boldsymbol{y}}$, and $\hat{\boldsymbol{z}}$. 
Without loss of generality, the BS array is assumed to be positioned at the height $h_r$ above the origin $O$, with its array aligned along the $x$-axis. Thus, the position of BS array center is given by $\mathbf{p}_r = (0, 0, h_r)$ and the position of the $r$-th receive antenna is given by 
$\mathbf{r}_{r} = (r^x_{r}, r^y_{r}, r^z_{r}) = (\delta_rd_{r}, 0, h_r)$, where $\delta_r = \frac{1}{2}(N_r + 1 - 2r)$ for $r = 1, 2, \dots, N_r$.
The UE array is assumed to be centered at $\mathbf{p}_t = (x_t, y_t, z_t)$. A local coordinate system $O'$-$x'y'z'$ is introduced, with its origin $O'$ fixed at the center of the UE array and its axes parallel to those of the global reference system. To characterize the orientation of the UE array antennas, the orientation unit vector $\mathbf{n}_t$ is given by
\begin{align}
\mathbf{n}_t =(n^x_{t}, n^y_{t}, n^z_{t})= (\sin \beta_t \sin \alpha_t, \sin \beta_t \cos \alpha_t, \cos \beta_t), \label{eq1}
\end{align}
where $\alpha_t$ is the tilt angle between the ${{x}}'$-axis and the array, i.e. the angle from the unit vector $\hat{\boldsymbol{y}}'$ to the projection of the orientation vector $\mathbf{n}_t$ onto the $x'O^{\prime}y'$ plane, and $\beta_t$ is the polar angle, i.e. the angle from the unit vector $\hat{\boldsymbol{z}}'$ to the orientation vector $\mathbf{n}_t$. $\mathbf{n}_t$ is orthogonal to the array axis. The position of the $t$-th transmit antenna element is given by
\begin{align}
\mathbf{u}_{t} \!=\! (u^x_{t}, u^y_{t}, u^z_{t})\!=\! \left( x_t \!+\!  \delta_t d_t \cos \alpha_t,\ y_t \!-\!\delta_t d_t \sin \alpha_t,\ z_t \right), \label{eq2}
\end{align}
where $\delta_{t} = \frac{1}{2}(N_t + 1 - 2 t)$, $t = 1, \dots, N_t$. The relative spatial relationship between the UE and BS array centers can be expressed as
\begin{align}
\mathbf{p}_t - \mathbf{p}_r  =
\begin{pmatrix}
\hat{\boldsymbol{x}} & \hat{\boldsymbol{y}} & \hat{\boldsymbol{z}}
\end{pmatrix}
\begin{pmatrix}
d_0 \cos \phi_r \sin \theta_r \\
d_0 \sin \phi_r \sin \theta_r \\
d_0 \cos \theta_r
\end{pmatrix}, \label{eq3}
\end{align}
where $d_0$ is the radial distance between the UE and BS array centers, $\phi_r \in[0,2 \pi)$ and $\theta_r \in[0, \pi]$ are the azimuth angle and elevation angle of the UE relative to the BS array center, respectively.
The distance $d_{t, r}$ between the $t$-th transmit antenna and the $r$-th receive antenna is given by:
\begin{align}
\!\!\!\!d_{t, r} & \!=\! \|\mathbf{r}_{r}\! - \!\mathbf{u}_{t}\| \notag \! =\! \Big[\left(r^r_{x}\! - \!u^t_{x}\right)^2 \!+\! \left(r^r_{y}\! - \!u^t_{y}\right)^2 \!+\! \left(r^r_{z} \!- \!u^t_{z}\right)^2\Big]^{\frac{1}{2}} \notag \\
 &\!= \Big[ d_0^2 + \delta_r^2 d_r^2 + \delta_t^2 d_t^2 - 2 \delta_r d_r d_0 \sin \theta_r \cos \phi_r \notag \\
 &\ \ \ \ -\! 2 \delta_r d_r \delta_t d_t \cos \alpha_t \!+\! 2 d_0 \delta_t d_t \sin \theta_r \cos(\phi_r \!+\! \alpha_t) \Big]^{\frac{1}{2}}\!\!\!. \label{eq4}
\end{align}

In the EM exposure scenario, we introduce a spherical head model of the user, centered at position $\mathbf{p_u}$ near the UE array.
During the uplink transmission, a portion of the radiated energy from the UE  is absorbed by the head.
Due to the limited penetration of mmWave into human tissue, energy deposition is predominantly concentrated on the surface. 
Thus, we spatially sample $M$ units on the surface of the head model for subsequent quantitative exposure assessment, indexed by $m = 1, 2, \ldots, M$.
In accordance with the IEEE International Committee on Electromagnetic Safety (ICES) guidelines, each sampling unit corresponds to a square tissue region with side length $a=2 \, \text{cm}$ \cite{r8}.  
The center point of the $m$-th sampling unit is given by the position vector $\mathbf{s}_m = (s^m_x, s^m_y, s^m_z)$.
The distance $d_{t, m}$ between the $t$-th transmit antenna and the $m$-th sampling point is given by
\begin{align}
d_{t, m} = \bigg[ & \left(s^m_x - \left(x_t + n_{t} \delta_t \cos \alpha_t\right)\right)^2 \notag \\
 & + \left(s^m_y - \left(y_t - n_{t} \delta_t \sin \alpha_t\right)\right)^2 + (s^m_z-z_t)^2 \bigg]^\frac{1}{2}. \label{eq5}
\end{align}

\subsection{Radiation Model}

\begin{figure}
\includegraphics[width=0.48\textwidth]{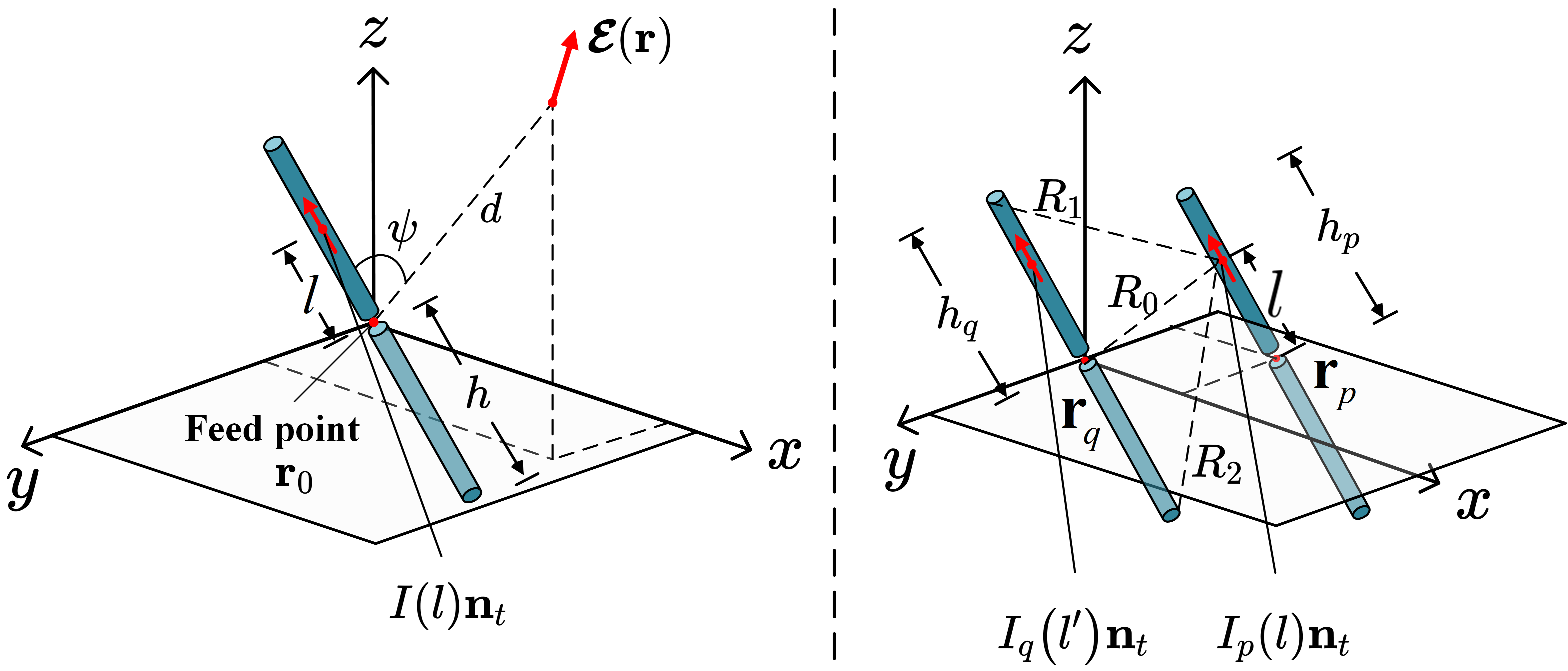}\vspace{-0.2cm}
\caption{Illustration of the EM propagation model (a) Single antenna; (b) Parallel array.}
\setlength{\parskip}{0.5cm plus4mm minus3mm}
\mbox{}
\label{f2}
\vspace{-0.8cm}
\end{figure}

From an EM perspective, antenna excitation originates from a feed voltage that induces a specific current distribution. 
As shown in Fig. \ref{f2}(a), we consider a linearly polarized dipole antenna located at position \(\mathbf{r}_0\) and oriented along the orientation unit vector \(\mathbf{n}_t\). Using the thin-wire approximation and assuming a sinusoidal current profile with the feed point at \(\mathbf{r}_0\), the current distribution along the antenna axis is \cite{rsinc}
\begin{align}
I(l) = I_0 \frac{\sin\left(k(h - |l|)\right)}{\sin (k h)}=I_m {\sin\left(k(h - |l|)\right)},
\label{eq6}
\end{align}
where \(l \in [-h, +h]\) is the coordinate along the antenna axis, $h$ is the half-length of the antenna, \({I}_{0}\) is the port current of the antenna ( time-harmonic factor \(e^{j\omega t}\) is omitted, where $\omega$ is the angular frequency), \({I}_{m}={{I}_{0}}/{{\sin (k h)}}\) is the maximum current and \(k = 2\pi/\lambda\) is the wavenumber.

For an arbitrary point $\mathbf{r}'$  along the antenna axis, the corresponding electric current density is given by \cite{ca}
\begin{align}
\mathbf{J}(\mathbf{r}') = I(l) \, \mathbf{n}_t \, \delta(\mathbf{r}' - \mathbf{r}_0 - l \mathbf{n}_t),\label{eq7}
\end{align} 
\noindent where  \(\delta(\cdot)\) is the Dirac delta function. The radiated electric field can then be derived from the EM theory, as summarized in the following lemma.

\begin{lemma}\label{aq1}(Single-Antenna Radiation Model): 
Consider the transmit antenna positioned at $\mathbf{r}_0$ and oriented along the orientation unit vector  $\mathbf{n}_t$. The electric field vector \(\boldsymbol{\mathcal{E}}(\mathbf{r})\) at an observation point $\mathbf{r}$ is given by
\begin{align}
\!\boldsymbol{\mathcal{E}}(\mathbf{r})\! =\!  \frac{{j \eta I_m}_{}}{2 \pi {d}}
\underbrace{\frac{\cos (k h \cos \psi) - \cos (k h)}{\sin \psi}}_{g(\hat{\mathbf{r}})}
\hat{\boldsymbol{\rho}}( \hat{\mathbf{r}}){e^{-j k d}},
\label{eq8}
\end{align}
where \(\eta \) is the free-space impedance, \(d = \|\mathbf{r}-\mathbf{r}_0\|\), \(\hat{\mathbf{r}} = (\mathbf{r}-\mathbf{r}_0)/d\) is the unit direction vector, \(\psi\) is the angle between the dipole axis \(\mathbf{n}_t\) and \(\hat{\mathbf{r}}\), i.e., \(\cos\psi = \mathbf{n}_t \cdot \hat{\mathbf{r}}\), $g(\hat{\mathbf{r}})$ is the normalized gain of the electric field, and $\hat{\boldsymbol{\rho}}( \hat{\mathbf{r}}) = \frac{ 1 }{\sin\psi}[(\mathbf{n}_t \cdot \hat{\mathbf{r}})\hat{\mathbf{r}} - \mathbf{n}_t]$ is the unit polarization vector. 
\end{lemma}
\begin{IEEEproof}
Please refer to Appendix A.    
\end{IEEEproof}

Building upon the single-antenna radiation model in Lemma \ref{aq1}, we generalize it to an array of parallel linear antennas.
When antennas of the array are positioned in proximity, their mutual coupling cannot be ignored \cite{rmuc}. This effect is quantified by the mutual impedance derived in the following lemma. 
\begin{lemma}\label{aq21}(Mutual Impedance): 
Consider the $p$-th and $q$-th antenna elements in a parallel array, centered at positions \(\mathbf{r}_p\) and \(\mathbf{r}_q\) with half-lengths \(h_p\) and \(h_q\), respectively, as illustrated in Fig.~\ref{f2}(b). The mutual impedance between them is given by
\begin{align}
Z_{pq} &\!= \!
\frac{j\eta}{4\pi \sin (k h_p) \sin (k h_q)} \!\int_{-h_p}^{h_p} F_{pq}(l)  \, dl,
\label{eq9}
\end{align}
and the kernel $F_{pq}(l)$ in \eqref{eq9} is defined as
\begin{align}
\!F_{pq}(l) \!=\! \left[\frac{e^{\small{-j k R_{1}}}}{R_{1}}\!+\!\frac{e^{\small{-j k R_{2}}}}{R_{2}}\!-\!2\! \cos (k h_q) \frac{e^{\small{-j k R_{0}}}}{R_{0}}\right]\!\sin\bigl[k(h_p \!-\! |l|)\bigr],
\notag
\end{align}
where $R_0 = \|\mathbf{r}_p-\mathbf{r}_q+l \mathbf{n}_t\|$, $R_1 = \|\mathbf{r}_p-\mathbf{r}_q+(l-h_q) \mathbf{n}_t\|$ and $R_2 = \|\mathbf{r}_p - \mathbf{r}_q + (l+h_q) \mathbf{n}_t\|$.
\end{lemma}
\begin{IEEEproof}
Please refer to Appendix B.    
\end{IEEEproof}

Similarly, the self-impedance \(Z_{pp}\) follows as a special case of Lemma \ref{aq21} with \(q = p\), with distances evaluated over the antenna surface. 
In this paper, we focus on half-wave dipoles.  For an \(N_t\)-element array, the relationship between the driving voltage vector \(\boldsymbol{v}_{t}\) and induced current vector \(\boldsymbol{i}\) is governed by the impedance matrix \(\mathbf{Z} \in \mathbb{C}^{N_t \times N_t}\):
\begin{align}
\underbrace{\begin{bmatrix} V_1 \\ V_2 \\ \vdots \\ V_{N_t} \end{bmatrix}}_{\Large{\boldsymbol{v}_{t}}} = \underbrace{\begin{bmatrix} Z_{11} & Z_{12} & \cdots & Z_{1N_t} \\ Z_{21} & Z_{22} & \cdots & Z_{2N_t} \\ \vdots & \vdots & \ddots & \vdots \\ Z_{N_t1} & Z_{N_t2} & \cdots & Z_{N_tN_t} \end{bmatrix}}_{\mathbf{Z}} \underbrace{\begin{bmatrix} I_1 \\ I_2 \\ \vdots \\ I_{N_t} \end{bmatrix}}_{\LARGE{\boldsymbol{i}}}, \label{eq10}
\end{align}
where \(\mathbf{Z}\) has entries $Z_{pq}$ given by Lemma \ref{aq21}. Note that the impedance matrix has non-conjugate transpose symmetry, as a consequence of the reciprocity relation $Z_{pq}=Z_{qp}$.

Consider the \(r\)-th receive antenna  located at position \(\mathbf{r}_{r}\). We define the array response vector  \(\mathbf{a}( \mathbf{r}_r) \in \mathbb{C}^{N_t \times 1}\) that captures the effect of the normalized gain and the spherical wavefront, which can be expressed as
\begin{align}
\mathbf{a}(\mathbf{r}_r) \!=\! \left[ g(\hat{\mathbf{r}}_{1,r})\frac{e^{-j k d_{1,r}}}{d_{1,r}},\; \ldots,\; g(\hat{\mathbf{r}}_{N_t,r})\frac{e^{-j k d_{N_t,r}}}{d_{N_t,r}} \right]^\mathsf{T},\label{eq11}
\end{align}
where  \(g(\cdot)\)  is the normalized gain given in \eqref{eq8}. According to \eqref{eq8}, \eqref{eq10}, and \eqref{eq11}, the amplitude vector of the electric field radiated by the array is
\begin{align}
\boldsymbol{\mathcal{A}}_{\text{arr}}(\mathbf{r}_r) = \frac{j\eta}{2\pi} \left( \mathbf{a}( \mathbf{r}_r) \odot ( \mathbf{Z}^{-1} \boldsymbol{v}_{t}) \right) ,\label{eq12}
\end{align}
By stacking the polarization vector from all transmit antennas into $\boldsymbol{\mathcal{P}}( \mathbf{r}_r) = \left[ \hat{\boldsymbol{\rho}}(\hat{\mathbf{r}}_{1,r}),\; \ldots,\; \hat{\boldsymbol{\rho}}(\hat{\mathbf{r}}_{N_t,r})\right]\in \mathbb{C}^{3 \times N_t}$, the total electric field vector at  the $r$-th receive antenna is given by
\begin{align}
\boldsymbol{\mathcal{E}}_{\text{tot}}( \mathbf{r}_r) = \boldsymbol{\mathcal{P}}( \mathbf{r}_r) \;   \boldsymbol{\mathcal{A}}_{\text{arr}}( \mathbf{r}_r) \;.
\label{eq13}
\end{align}
The expressions \eqref{eq12} and \eqref{eq13} provide a mathematically tractable model for channel modeling and EM exposure assessment.

\subsection{Channel Model}
Building upon the radiation model presented in the preceding section, a mathematical framework is established to characterize the EM wave propagation between the UE and BS. This section presents an electromagnetically consistent channel model that encapsulates this EM propagation process.

At the transmitter side, the transmitted physical signal is characterized by the voltage that excites the transmit antenna. Let the excitation voltage vector be \(\boldsymbol{v}_{t} = v_0 s_t\mathbf{w}\), where \(v_0\) is a power-normalization coefficient, $s_t$ is the unit-power data symbol and \(\mathbf{w} \in \mathbb{C}^{N_t\times1}\) is the beamforming vector containing the complex excitation weights for each transmit antenna. Without loss of generality, we assume a constant envelope modulation scheme, i.e.  \(|s_t|^2 = 1\). Then, the current vector follows from Ohm’s law as \(\boldsymbol{i} = \mathbf{Z}^{-1} \boldsymbol{v}_{t} = v_0 s_t\mathbf{Z}^{-1} \mathbf{w}\). The average transmit power of the array is given by
\begin{align}
P(\mathbf{w})=\frac{1}{2}\operatorname{Re}\left(\boldsymbol{v}_{t}^\mathsf{H} \boldsymbol{i}\right)=\frac{v_0^2}{2} \operatorname{Re}\left(\mathbf{w}^\mathsf{H} \mathbf{Z}^{-1} \mathbf{w}\right).
\label{121}
\end{align}
Since the impedance matrix is symmetric, i.e., \(\mathbf{Z} = \mathbf{Z}^\mathsf{T}\), the Rayleigh–Ritz theorem provides an upper bound of \eqref{121} as
\begin{align}
P(\mathbf{w}) \leq \frac{v_0^2}{2} \lambda_{\max }\left(\operatorname{Re}\left(\mathbf{Z}^{-1}\right)\right)\|\mathbf{w}\|^2, 
\label{123}
\end{align}
where \(\lambda_{\max}(\cdot)\) denotes the largest eigenvalue of its argument. 
To ensure that the power constraint \(P(\mathbf{w}) \le P_{\text{max}}\) holds for any admissible beamforming vector \(\mathbf{w}\), we set the normalization factor as $
v_0 = \small{\sqrt{{2}/\lambda_{\max}\!\bigl(\operatorname{Re} \bigl( \mathbf{Z}^{-1} \bigr)\bigr)}}
$. Consequently, the power constraint is reduced to a simple norm condition on the beamforming vector $\mathbf{w}$, given as
\begin{align}
\|\mathbf{w}\|^2 \le P_{\text{max}}.
\end{align}
Then, the equivalent channel is derived in Proposition \ref{p11}
\begin{proposition}(Equivalent Channel Vector)\label{p11}:
Consider the transmission from the UE array to the $r$-th receive antenna. The equivalent EM  channel vector \(\mathbf{h}_{r} \in \mathbb{C}^{1 \times N_t}\) is given by
\begin{align}
\mathbf{h}_{r} = \frac{j\eta A_{F}v_0}{2\pi} \; \boldsymbol{\mu}_{r} \; \operatorname{diag}\big( \mathbf{a}(\mathbf{r}_{r}) \big) \; \mathbf{Z}^{-1},
\label{eq:channel_vector}
\end{align}    
where \(A_{F}\)  is an antenna factor converting field amplitude to voltage \cite{r23} and the \(\boldsymbol{\mu}_{r} \in \mathbb{C}^{1 \times N_t}\)  is the polarization mismatch vector.
\end{proposition}
\begin{IEEEproof}
At the receiver side, the voltage induced on each receive antenna depends on its polarization alignment with the incident EM wave. For a linearly polarized antenna oriented along \(\hat{\boldsymbol{\rho}}_r\), we denote the polarization mismatch vector \(\boldsymbol{\mu}_{r} \in \mathbb{C}^{1 \times N_t}\) as the projection coefficients of the receive antenna’s polarization onto the polarization of the wave arriving from transmit antennas, which can be expressed as
\begin{align}
\boldsymbol{\mu}_{r}= |\hat{\boldsymbol{\rho}}_r^\mathsf{T} \boldsymbol{\mathcal{P}}(\mathbf{r}_{r})| = \left[ |\hat{\boldsymbol{\rho}}_r \cdot \hat{\boldsymbol{\rho}}(\hat{\mathbf{r}}_{1,r})|, \dots,| \hat{\boldsymbol{\rho}}_r \cdot \hat{\boldsymbol{\rho}}(\hat{\mathbf{r}}_{N_t,r})| \right]. \notag
\label{eq:polarization_mismatch_vector}
\end{align}

When the BS is distant from the UE array, the direction vectors from each transmit antenna to the BS are approximately parallel, i.e., \(\hat{\mathbf{r}}_{1,r} \approx \cdots \approx \hat{\mathbf{r}}_{N_t,r} \triangleq \hat{\mathbf{r}}_r\). Thus, the mismatch vector  simplifies to $\boldsymbol{\mu}_{r} \approx |\hat{\boldsymbol{\rho}}_r \cdot \hat{\boldsymbol{\rho}}(\hat{\mathbf{r}}_r, \mathbf{n}_t)| \mathbf{1}_{N_t}$. Consequently, the induced voltage \(V_{r}\)  at the $r$-th receive antenna can be written as
\begin{align}
{V}_{r} (\mathbf{r}_{r})
&= A_{F} \; \boldsymbol{\mu}_{r} \;  \boldsymbol{\mathcal{A}}_{\text{arr}}(\mathbf{r}_{r})\; \notag \\
&= \underbrace{\frac{j\eta A_{F}v_0}{2\pi} \; \boldsymbol{\mu}_{r} \; \operatorname{diag}\big( \mathbf{a}(\mathbf{r}_{r}) \big) \; \mathbf{Z}^{-1}}_{\,\mathbf{h}_{r}} \; \mathbf{w}s_t, 
\end{align}
where the channel vector $\mathbf{h}_{r} $ encapsulates the mapping from the transmit symbol to the voltage induced at the receiver. This formulation provides a physically comprehensive model that incorporates spherical‑wave propagation, antenna radiation gain, polarization mismatch, and mutual coupling effects.
\end{IEEEproof}

We extend the model in \eqref{eq:channel_vector} to the complete channel matrix   \(\mathbf{H} \in \mathbb{C}^{N_r \times N_t}\). The channel matrix between the UE and BS can be obtained as
\begin{align}
\mathbf{H} = \begin{bmatrix} \mathbf{h}_{1}^{ T }, \mathbf{h}_{2}^{ T }, \dots ,\mathbf{h}_{N_r}^{ T } \end{bmatrix}^{ T }.
\label{eq:MIMO_channel_matrix}
\end{align}

Consequently, the singal vector \(\mathbf{y} \in \mathbb{C}^{N_r \times 1}\) received by the BS is given by
\begin{align}
\mathbf{y} = \mathbf{H} \mathbf{w}{s}_t + \mathbf{n}, 
\end{align}
where $\mathbf{n} \sim \mathcal{C N}\left(\mathbf{0}, \sigma^2\mathbf{I}_{N_r}\right)$ represents additive white Gaussian noise.

\section{EM Exposure Constraints}\label{Section3}
This section develops an integrated analytical framework for thermal exposure assessment in mmWave communication systems. We first introduce the fundamental bioheat transfer model that governs EM heating in tissue. Then, we extend it to the dynamic scenario where incident PD varies over time. A closed-form thermal impulse response model is derived to capture the thermal inertia of tissue. Finally, leveraging this dynamic model, we formulate an exposure-aware beamforming optimization problem with a long-term thermal budget constraint.
\subsection{Thermal Model}\label{SECTION3a}
International safety guidelines for mmWave frequencies are mainly designed to mitigate excessive tissue heating induced by EM exposure. To mechanistically assess the thermal response, we employ the BHTE as \cite{rBHTE}
\begin{align}
\!\!\kappa \nabla^2 T(\mathbf{r}, t) \!-\! \rho^2 C_p \omega_b T(\mathbf{r}, t) \!+\! \rho \mathrm{SAR}(\mathbf{r}) \!=\! \rho C_p \frac{\partial T(\mathbf{r}, t)}{\partial t}, \label{BHTE1}
\end{align}
where \(T(\mathbf{r}, t)\) denotes the temperature rise above the baseline (pre-exposure) level, \(\omega_b\) is the blood perfusion rate, \(\kappa\) is the thermal conductivity, \(C_p\) is the specific heat capacity, and \(\rho\) is the tissue density \cite{r24}(parameters are summarized in Table~\ref{tab:params}).

In \eqref{BHTE1}, the heat deposited by EM exposure is modeled as the external source term \(\rho\,\mathrm{SAR}(\mathbf{r})\). The SAR quantifies the absorbed power per unit mass of tissue and is expressed as \cite{r22}
\begin{align}
\mathrm{SAR}(\mathbf{r}) = \frac{\mathcal{I}(\mathbf{r})\, T_{\mathrm{tr}}}{\rho \delta_p} e^{-\frac{z}{\delta_p}},
\end{align}
where \(\mathcal{I}(\mathbf{r})\) is the incident PD on the tissue surface, \(T_{\mathrm{tr}}\) is the power transmission coefficient of the skin, \(\delta_p\) is the power penetration depth, and \(z\) is the depth coordinate normal to the surface. In the mmWave band, \(\delta_p\) is on the order of millimeters or less. Thus, taking the analytical limit \(\delta_p \to 0\), the heat source reduces to
\begin{align}
\rho \, \mathrm{SAR}(\mathbf{r}) \;\xrightarrow[]{\delta_p \to 0}\; T_{\mathrm{tr}} \, \mathcal{I}(\mathbf{r}) \, \delta(z). \label{eq:surface_heat}
\end{align}
Physically, \eqref{eq:surface_heat} confirms that mmWave energy is absorbed predominantly in a superficial layer, and the heating can be directly linked to the incident PD.

The BHTE \eqref{BHTE1} reveals that the surface temperature rise is ultimately limited by the balance between heat deposition and convective cooling by blood flow. The process is characterized by two intrinsic scales:
\begin{align}
\tau_\text{{th}} = \frac{1}{\omega_b \rho} \approx 500\ \text{s}, \qquad
R_\text{{th}} = \sqrt{\frac{\kappa}{\rho^2 C_p \omega_b}} \approx 7\ \text{mm}. \label{eq:scales}
\end{align}
Here, \(\tau_\text{{th}}\) represents the characteristic time for heat removal via blood perfusion, and \(R_\text{{th}}\) denotes the spatial scale over which perfusion dominates thermal regulation.   Collectively, these scaling parameters delineate the spatiotemporal dynamics of the temperature rise
under EM exposure.

\begin{table}[!t]
\centering
\caption{Thermal parameters of human skin tissue.}\label{tab:params}
\begin{tabular}{lcll}
\toprule
Symbol & Value & Unit & Physical Meaning \\
\midrule
\(\kappa\) & 0.37 & \(\mathrm{W/(m\cdot{}^{\circ}C)}\) & Thermal conductivity \\
\(\rho\)  & 1109 & \(\mathrm{kg/m^{3}}\) & Density \\
\(C_{p}\) & 3390 & \(\mathrm{J/(kg\cdot{}^{\circ}C)}\) & Specific heat capacity \\
\(\omega_b\) & 106 & \(\mathrm{m\ell/(min\cdot kg)}\) & Blood perfusion rate \\
\bottomrule
\end{tabular}
\end{table}

\subsection{Thermal Response to Time-Varying Exposure}
The conventional model presented in the previous subsection cannot capture dynamic variations in incident power density due to UE mobility and beamforming, which presents a critical challenge to accurate real-time thermal safety assessment. To address this, a thermal impulse response model that characterizes the dynamic bioheat transfer process is developed below.

We discretize the exposure timeline into slots of duration \(\Delta t\), indexed by \(n = 1, 2, \dots, N\). The UE array’s position and orientation vary across different slots. Denote \(\mathcal{I}_m[n]\) as the incident PD at the \(m\)-th location \(\mathbf{s}_m\) during the \(n\)-th slot, which is derived as 
\begin{align}
\mathcal{I}_m[n] &= \frac{\|\boldsymbol{\mathcal{E}}_{\text{tot},m}[n]\|^2}{2 \eta} \notag \\
&= \frac{1}{2 \eta}
\Big{\|}
\underbrace{\frac{j \eta v_0}{2 \pi} \boldsymbol{\mathcal{P}}_m[n]
\operatorname{diag}\left(\mathbf{a}_m[n]\right) \mathbf{Z}^{-1}}
_{\Phi_{m}[n]} \mathbf{w}[n]\Big{\|}^2,
\label{eq:PD_calculation}
\end{align}
where the subscript \(m\) indicates the location \(\mathbf{s}_m\) for brevity, and \(\Phi_m[n] \in \mathbb{C}^{3 \times N_t}\) is denoted as the exposure array manifold at slot \(n\), encapsulating the combined effect of polarization, spherical wave propagation, and mutual coupling from the transmit array to \(\mathbf{s}_m\). Then, the incident PD can be compactly expressed in quadratic form as
\begin{align}
\mathcal{I}_m[n] = \frac{1}{2\eta} \, \mathbf{w}^\mathsf{H}[n] \, \Phi_m^\mathsf{H}[n] \, \Phi_m[n] \, \mathbf{w}[n], 
\label{eq:PD_quadratic}
\end{align}
which explicitly shows that the temporal variation of the exposure level originates from two factors, i.e., the time-varying array manifold \(\Phi_m[n]\) that captures UE mobility, and the adaptive beamforming vector \(\mathbf{w}[n]\) governed by the transmission strategy.

To characterize the tissue’s dynamic thermal response under such time-varying PD, we derive a fundamental thermal impulse response model via the Green’s function approach, formalized in Proposition~\ref{prop:impulse_response}.

\begin{proposition}[Thermal Impulse Response]
\label{prop:impulse_response}
For a finite square irradiation area \(D_m = [-a, a] \times [-a, a]\) on the tissue surface, the thermal impulse response \(g_m(t)\) at a sampling point \(\mathbf{s}_m\) is given by
\begin{align}
g_m(t)
= \frac{T_{\mathrm{tr}} R_\text{{th}}}{\kappa}\,
\frac{1}{\sqrt{\pi\tau_\text{{th}} t}}\,
e^{-t/\tau_\text{{th}}}, \qquad t > 0,
\end{align}
where \(R_\text{{th}}\) and \(\tau_\text{{th}}\) are the characteristic length and time scales defined in \eqref{eq:scales}.
\end{proposition}

\begin{IEEEproof}
Consider the semi-infinite tissue with an adiabatic boundary condition at the skin surface. The temperature rise due to a surface heat source can be expressed via convolution with the Green’s function
\begin{align}
\!T(\mathbf{r}, t) \!=\! \int_0^t \!\iint_{D} \!\underbrace{G\left( \left| \mathbf{r} - \mathbf{r}' \right|, t - t' \right)}_{\text{Green's\ function}}\! \cdot \!\underbrace{\frac{T_{\mathrm{tr}} \mathcal{I}(\mathbf{r})}{\rho C_p}}_{\text{Heat\ source}}  dx'  dy'  dt', \label{eq:convolution_integral}
\end{align}
 and the Green's function for the BHTE is \cite{rgreen}
\begin{align}
G\left(\mathbf{r} - \mathbf{r}' , t - t' \right) \!=\! \frac{2}{\left[4\pi \alpha (t \!- \!t')\right]^\frac{3}{2}} \exp\!\left( -\frac{\left| \mathbf{r} - \mathbf{r}' \right|^2}{4\alpha (t \!-\! t')}\! -\! \frac{t\! -\! t'}{\tau_\text{{th}}}\right), \notag
\end{align}
where thermal diffusivity \(\alpha = \kappa/(\rho C_p)\), and the factor 2  arises from the image source. accounts for the unidirectional flow of heat into the surface. For a unit-step EM exposure \(\mathcal{I}(\mathbf{r}') = u(t)\) over \(D_m\), where \(u(t)\) is the Heaviside step function, evaluating \eqref{eq:convolution_integral} at \(\mathbf{s}_m\) gives
\begin{align}
T\!\left({\mathbf{s}}_m, t\right) & \!=\!\!\int_0^t\! \frac{2 T_{\operatorname{tr}}}{\rho C_p(4 \pi \alpha u)^{\frac{3}{2}}}\!\left[\iint_{D_m} \!\!e^{\left(-\frac{x^{\prime 2}+y^{\prime 2}}{4 \alpha u}\right)} d x^{\prime} d y^{\prime}\right] \!e^{\frac{-u}{\tau_\text{{th}}}} d u\notag \\
& \!=\!\frac{2 T_{\operatorname{tr}}}{\rho C_p(4 \pi \alpha)^{1 / 2}} \int_0^t \frac{e^{\frac{-u}{\tau_\text{{th}}}}}{\sqrt{u}}\left[\operatorname{erf}\left(\frac{a}{\sqrt{4 \alpha u}}\right)\right]^2 d u
\label{eq:temp_integral}
\end{align}
where \(u = t - t'\). For typical exposure averaging intervals \(t < 6\ \text{min}\) and the sampling unit side length \(a=2\text{cm}\), the approximation \(\operatorname{erf}\!\bigl(a/\sqrt{4\alpha u}\bigr) \approx 1\) holds. Substituting this into \eqref{eq:temp_integral} yields
\begin{align}
T(\mathbf{s}_m, t)
& \approx \frac{2 T_{\mathrm{tr}}}{\rho C_p (4\pi \alpha)^{1/2}} \int_0^t \frac{e^{-u / \tau_\text{{th}}}}{\sqrt{u}} \, du \notag \\
& = \frac{ T_{\mathrm{tr}} R_\text{{th}}}{\kappa} \operatorname{erf}\left( \sqrt{\frac{t}{\tau_\text{{th}}}} \right).
\label{eq:temp_rise_step}
\end{align}
The thermal impulse response is defined as the time derivative of the temperature rise, given by
\begin{align}
g_m(t) = \frac{d}{dt} T(\mathbf{s}_m, t) = \frac{T_{\mathrm{tr}} R_\text{{th}}}{\kappa}\, \frac{1}{\sqrt{\pi\tau_\text{{th}} t}}\, e^{-t/\tau_\text{{th}}}.
\end{align}
This completes the proof.
\end{IEEEproof}

Using the incident PD in \eqref{eq:PD_calculation} and the impulse response from Proposition~\ref{prop:impulse_response}, the temperature rise at discrete time \(n\) can be explicitly calculated as
\begin{align}
T_m[n] &\approx \sum_{k=1}^{n} \mathcal{I}_{m}[k] \int_{(k-1)\Delta t}^{k\Delta t} g_m(n\Delta t - \xi)  \, d\xi \notag \\
&= \frac{T_{\mathrm{tr}} R_\text{{th}}}{\kappa} \sum_{k=1}^{n} \mathcal{I}_{m}[k] \, \xi_{n-k},
\label{36a}
\end{align}
where the coefficients \(\xi_{i} = \operatorname{erf}(\sqrt{\frac{(i+1)\Delta t}{\tau_\text{{th}}}}) - \operatorname{erf}(\sqrt{\frac{i\Delta t}{\tau_\text{{th}}}})\) are pre-computable. For small $\Delta t$ and large $n$, the sequence \(\{\xi_i\}\) decays approximately as \(\xi_i \approx \xi_{i-1} e^{-\Delta t/\tau_\text{{th}}}\). Consequently, \eqref{36a} reduces to a first‑order Markov recursion:
\begin{equation}
T_m[n]=e^{-\Delta t / \tau_\text{{th}}} T_m[n-1] + \frac{T_{\mathrm{tr}} R_\text{{th}}}{\kappa}\xi_0 \mathcal{I}_{m}[n]
\label{37a}
\end{equation}

This recursive form captures the essential inertia of the bioheat process, where the temperature \(T_m[n]\) depends only on its previous state \(T_m[n-1]\) and the current incident PD \(\mathcal{I}_m[n]\). The coefficient \(e^{-\Delta t/\tau_\text{{th}}}\) quantifies the heat retained due to thermal inertia. Importantly, it reveals that if the incremental temperature rise is kept below the heat dissipation, the temperature can be stabilized within a safe range. This inherent thermal-budget insight establishes the foundation for the subsequent optimization framework.

\subsection{Exposure-Aware Beamforming Problem}
Based on the channel model in \eqref{eq:MIMO_channel_matrix} and the exposure model in \eqref{36a}, we can formulate an adaptive exposure-aware beamforming problem. It aims to maximize the average received SNR over \(N\) time slots by optimizing beamforming vectors \(\{\mathbf{w}[n]\}_{n=1}^N\), subject to transmit power constraint and long‑term temperature constraints.

In contrast to conventional conservative approaches that enforce instantaneous exposure limits, the proposed framework exploits the thermal model in \eqref{36a} and constructs a long-term exposure constraint at each sampling point:
\[
\frac{1}{N} \sum_{n=1}^{N} T_m[n] \leq T_{\text{th}}, \qquad \forall m \in \{1,\dots,M\},
\]
where \(T_{\text{th}}\) is a predefined temperature safety threshold. This formulation effectively converts rigid instantaneous exposure limits into a flexible long-term thermal budget that can be allocated intelligently across time. By satisfying the constraint at every sampling point, the framework inherently controls the spatial peak temperature, thereby ensuring compliance with safety guidelines.

Then, the long-term exposure-aware optimization problem is stated as 
\begin{align}
(\mathcal{P}1):\quad & \max_{\{\mathbf{w}[n]\}_{n=1}^N} \; \frac{1}{N} \sum_{n=1}^N \|\mathbf{H}[n]\mathbf{w}[n]\|^2 \\[4pt]
\text{s.t.} \quad & \|\mathbf{w}[n]\|^2 \leq P_{\text{max}}, \qquad \forall n \in \{1,\dots,N\}, \\[2pt]
& \frac{1}{N} \sum_{n=1}^N T_m[n] \leq T_{\text{th}}, \qquad \forall m \in \{1,\dots,M\}, \label{temcon}
\end{align}
where $P_{\text{max}}$ denotes the maximum transmit power constraint and the temperature rise \(T_m[n]\) at each sampling point is governed by  \eqref{36a}.

Solving problem \((\mathcal{P}1)\) is challenging. First, its objective and constraints involve long-term time averages, which preclude direct solution by conventional optimization techniques that typically require instantaneous or deterministic problem formulations. Second, in practice, the inherent randomness of wireless channel fading and relative transceiver motion necessitates a causal decision-making policy. Specifically, the beamforming vector \(\mathbf{w}[n]\) at each time slot \(n\) must be determined based solely on current and past observations, without knowledge of future channel states or future exposure conditions. To address these, we leverage the Lyapunov optimization framework to obtain low-complexity online solutions for this problem in the next section.

\section{Exposure-Aware Beamforming Design}\label{Section4}
In this section, we employ the Lyapunov optimization framework. This approach enforces the long-term temperature constraints by ensuring the stability of virtual queues, thereby transforming the original long-term problem ($\mathcal{P}$1) into a sequence of deterministic per-slot optimization problems.

\subsection{Lyapunov Optimization Formulation}
We first convert the long-term averaging constraints into a queue stability constraint. For each sampling point \( m \in \{1,2,\ldots,M\} \),  we define a virtual queue that evolves as
\begin{align}
Q_m[n+1] = \max\left\{0, Q_m[n] + T_m[n] - T_{\text{th}}\right\}.
\label{eq:virtual_queue}
\end{align}
Note that the queue of each point is initially empty, i.e., $Q_m[1]=0, \forall m \in \{1,\dots,M\}$.
The virtual queue quantifies a measure of the cumulative constraint violation over past slots.
\begin{lemma}[Constraint Satisfaction via Queue Stability] \label{lem:constraint_satisfaction}
If the virtual queues \( \{Q_m[n]\}_{m=1}^M \) are rate stable, i.e., $\lim_{n \to \infty} \left[{Q_m[n]}\right]/{n} = 0 ,\, \forall m$, the temperature constraint is satisfied: 
\begin{align}
\limsup_{N \to \infty} \frac{1}{N} \sum_{n=1}^{N} T_m[n] \leq T_{\text{th}}, \quad \forall m.
\end{align}
\end{lemma}

\begin{IEEEproof}
According to
 \eqref{eq:virtual_queue}, we have
\begin{align}
Q_m[n+1] &= \max\left\{0, Q_m[n] + T_m[n] - T_{\text{th}}\right\} \notag \\
&\geq Q_m[n] + T_m[n] - T_{\text{th}}, \label{eq:queue_inequality}
\end{align}
which holds because \( \max\{0, a\} \geq a \) for any \( a \geq 0\).
Rearranging \eqref{eq:queue_inequality} and summing over \( n = 1, 2, \dots, N \) yields
\begin{align}
\frac{1}{N} \sum_{n=1}^{N} T_m[n] - T_{\text{th}} \leq \frac{Q_m[N+1] - Q_m[1]}{N}. \label{eq:temp_bound}
\end{align}
Applying the $\limsup$ as \( N \to \infty \), we have
\begin{align}
\limsup_{N \to \infty}\!  \frac{1}{N} \!\sum_{n=1}^{N} T_m[n]\!-\! T_{\text{th}} 
&\leq \lim_{N \to \infty} \frac{Q_m[N+1] }{N}. \label{eq:expect_bound}
\end{align}
By the assumption of the mean rate stability, i.e., \( \limsup_{n \to \infty} [Q_m[n]]/n = 0 \), we have 
\begin{align}
\lim_{N \to \infty} \frac{1}{N} \sum_{n=1}^{N} T_m[n] \leq T_{\text{th}}.
\end{align}
This completes the proof.
\end{IEEEproof}

Subsequently, to ensure the stability of the virtual queues, we define a quadratic Lyapunov function as
\begin{align}
L(\mathbf{Q}[n]) = \frac{1}{2} \sum_{m=1}^M Q_m^2[n],
\label{q12}
\end{align}
where we define $\mathbf{Q}[n] = [Q_1[n], Q_2[n], \ldots, Q_M[n]]^\mathsf{T}$ to represent the queue lengths of all sampling points at the time slot $n$. The one-step conditional Lyapunov drift measures the expected growth of the Lyapunov function from one slot to the next, defined as \cite{rlyap1}
\begin{align}
\Delta(\mathbf{Q}[n]) = \mathbb{E}\left[L(\mathbf{Q}[n+1]) - L(\mathbf{Q}[n]) \mid \mathbf{Q}[n]\right].
\end{align}

To stabilize the virtual temperature queues, we can obtain an optimal scheme that minimizes a bound of the drift-plus-penalty expression at each slot $t$, which is defined as
\begin{align}
\Delta(\mathbf{Q}[n]) - V\mathbb{E}\left[\|\mathbf{H}[n]\mathbf{w}[n]\|^2 \mid \mathbf{Q}[n]\right], 
\label{44a}
\end{align}
where $V \geq 0$ is a control parameter that adjusts the trade‑off between constraint satisfaction and reward maximization. A larger $V$ places more emphasis on maximizing the system performance, while a smaller $V$ prioritizes stricter adherence to the exposure constraint.

Directly minimizing \eqref{44a} is intractable since the non-linearity is introduced by the max function in the drift term. Instead, we derive a deterministic upper bound that can be optimized causally. First, consider the change in the quadratic Lyapunov function for each virtual queue:
\begin{align}
Q&_m^2[n+1] - Q_m^2[n] \notag\\
&= \left(\max\left\{0, Q_m[n] + T_m[n] - {T}_{\mathrm{th}}\right\}\right)^2 - Q_m^2[n] \notag \\
&\leq \left(Q_m[n] + T_m[n] - {T}_{\mathrm{th}}\right)^2 - Q_m^2[n] \notag\\
&= \left(T_m[n] - {T}_{\mathrm{th}}\right)^2 + 2Q_m[n]\left(T_m[n] - {T}_{\mathrm{th}}\right)\notag \notag\\
&\leq (T_{\max} + {T}_{\mathrm{th}})^2 + 2Q_m[n]\left(T_m[n] - {T}_{\mathrm{th}}\right),
\label{eq:queue_diff_bound}
\end{align}
where the inequality follows from $\{\max\{a,0\}\}^2 \leq a^2$ and the temperature rise $T_m[n]$ is bounded by $T_{\max}$ for all $m$ and $n$ and we have $\left( T_m[n] - T_{\text{th}} \right)^2 \leq (T_{\max} + T_{\text{th}})^2$.

Summing \eqref{eq:queue_diff_bound} over all $m$ and taking the conditional expectation yields an upper bound as
\begin{align}
\Delta(\mathbf{Q}[&n]) - V\mathbb{E}\left[\|\mathbf{H}[n]\mathbf{w}[n]\|^2 \mid \mathbf{Q}[n]\right] \notag\\
&=\mathbb{E}\left[L(\mathbf{Q}[n+1]) - L(\mathbf{Q}[n]) \mid \mathbf{Q}[n]\right]\notag\\
&\qquad - V\mathbb{E}\left[\|\mathbf{H}[n]\mathbf{w}[n]\|^2 \mid \mathbf{Q}[n]\right]\notag\\
&\leq B + \sum_{m=1}^M Q_m[n] \cdot \mathbb{E}\left[T_m[n] - {T}_{\mathrm{th}} \mid \mathbf{Q}[n]\right] \notag \\
&\qquad- V\mathbb{E}\left[\|\mathbf{H}[n]\mathbf{w}[n]\|^2 \mid \mathbf{Q}[n]\right], 
\label{46a}
\end{align}
where $B = \frac{M}{2} (T_{\max} + T_{\text{th}})^2$ is a finite constant. Following the drift-plus-penalty framework, we opportunistically minimize the expectation to optimize the upper bound of the drift-plus-penalty function. Since constant terms do not affect the solution, we derive the per-slot optimization problem by minimizing the instantaneous version of the upper bound, which can be reformulated as
\begin{align}
(\mathcal{P}2):\min_{\mathbf{w}[n]} \quad & \sum_{m=1}^M Q_m[n] \cdot T_m[n] - V \|\mathbf{H}[n]\mathbf{w}[n]\|^2 \notag\\
&\text{s.t.} \quad  \|\mathbf{w}[n]\|^2 \leq P_{\text{max}}.\notag 
\end{align}

Note that only the current term in the convolution affects the immediate decision. Substituting the thermal dynamics from \eqref{37a}, we obtain the final per‑slot optimization problem as
\begin{align}
(\mathcal{P}3):\max_{\mathbf{w}[n]} \,\,  V \|\mathbf{H}[n]\mathbf{w}&[n]\|^2 \!-\! \frac{\xi_0}{2\eta} \!\sum_{m=1}^{M} Q_m[n] \, \|\boldsymbol{\Phi}_m[n]\mathbf{w}[n]\|^2 \notag \\ 
\text{s.t.} \quad & \|\mathbf{w}[n]\|^2 \leq P_{\text{max}}.\notag 
\end{align}

The per‑slot optimization problem ($\mathcal{P}3$) captures a fundamental trade-off between the system performance and the accumulated exposure constraint.  It can be reduced to a quadratically constrained quadratic program (QCQP): 
\begin{align}
\max_{\mathbf{w}[n]} \quad & \mathbf{w}^\mathsf{H}[n] \, \mathbf{A}[n] \, \mathbf{w}[n] \label{eq48}
\end{align}
where $
\mathbf{A}[n]=V \mathbf{H}^\mathsf{H}[n] \mathbf{H}\![n]-\frac{\xi_0}{2 \eta} \sum_{m=1}^M \!Q_m[n]\boldsymbol{\Phi}_m^\mathsf{H}[n] \boldsymbol{\Phi}_m\![n]$.
Since \(\mathbf{A}[n]\) is Hermitian and the power constraint $\|\mathbf{w}[n]\|^2 \leq P_{\text{max}}$ defines a spherical feasible set, the QCQP admits a closed‑form optimal solution, given by
\begin{align}
\mathbf{w}^*[n] = 
\begin{cases}
\sqrt{P_{\text{max}}} \; \mathbf{v}_{\text{max}}[n], & \lambda_{\max}\bigl(\mathbf{A}[n]\bigr) > 0, \\[4pt]
\mathbf{0}, & \text{otherwise},
\end{cases}
\label{eq:optimal_solution_mimo}
\end{align}
where \(\mathbf{v}_{\max}[n]\) denotes the unit‑norm eigenvector corresponding to the largest eigenvalue \(\lambda_{\max}\bigl(\mathbf{A}[n]\bigr)\). The second case corresponds to a safety-oriented silent mode where transmission is suspended to prevent further temperature rise.

Based on this closed‑form solution, we propose Algorithm \ref{alg:exposure_aware_beamforming} for adaptive exposure-aware beamforming to solve the original long-term optimization problem $(\mathcal{P}1)$. At each slot, the algorithm computes the optimal beamforming vector based on the current channel and exposure states, updates the temperature rise via \eqref{36a}, and then updates the virtual queues according to \eqref{eq:virtual_queue}. The control parameter $V$ allows the system to trade between communication performance and exposure safety. For any finite $V$, the algorithm guarantees stability of the virtual queues and convergence to a near-optimal solution of the original long-term problem. This is shown below.

\begin{algorithm}[t]
\caption{Adaptive Exposure-Aware Beamforming via Lyapunov Optimization}
\label{alg:exposure_aware_beamforming}
\begin{algorithmic}[1]
\STATE \textbf{Initialization}
\STATE Set control parameter $V \geq 0$ and let $n = 1$
\STATE Set initial virtual queues $Q_m[1] = 0$ for all $m$
\WHILE{$n \leq N$}
    \STATE Compute channel matrix $\mathbf{H}[n]$ according to \eqref{eq:MIMO_channel_matrix}; 
    \STATE Compute exposure array manifold $\{\boldsymbol{\Phi}_m[n]\}_{m=1}^M$ 
  \\according to \eqref{eq:PD_calculation};
    \STATE Obtain $\mathbf{A}[n]$ according to \eqref{eq48};
    \STATE Obtain the largest eigenvalue $\lambda_{\max}[n]$  and corresponding unit eigenvector $\mathbf{v}_{\max}[n]$
    \IF{$\lambda_{\max}[n] > 0$}
        \STATE $\mathbf{w}^*[n] = \sqrt{P_{\text{max}}} \cdot \mathbf{v}_{\max}[n]$ 
    \ELSE
        \STATE $\mathbf{w}^*[n] = \mathbf{0}$   (Safety-oriented silent mode)
    \ENDIF
    \STATE \textbf{Update system state:}
    \STATE Apply $\mathbf{w}^*[n]$ and update current temperature rise $\{T_m[n]\}_{m=1}^M$ according to \eqref{36a}
    \STATE Update all virtual queues $Q_m[n+1] = \max\left\{0, Q_m[n] + T_m[n] - T_{\text{th}}\right\}$, $\forall m$;
    \STATE Update $n = n + 1$;
\ENDWHILE 
\end{algorithmic}
\end{algorithm}

\subsection{Algorithm Performance Analysis}
The following proposition reveals that the algorithm can approach the optimal time-average received power while satisfying the long-term temperature constraints, with a provable performance bound controlled by the parameter $V$.

\begin{proposition}\label{Overall Performance Bound} (Performance Bound):
For any control parameter \(V \geq 0\), the proposed algorithm achieves the following performance bound:
\begin{align}
\lim_{N \to \infty} \frac{1}{N} \sum_{n=1}^{N} \mathbb{E}\bigl[ \|\mathbf{H}[n]\mathbf{w}[n]\|^2 \bigr] \; \geq \; \mathcal{G}_{\mathrm{opt}} - \frac{B}{V},
\end{align}
where \(\mathcal{G}_{\mathrm{opt}}\) denotes the optimal value of problem \((\mathcal{P}1)\), and \(B\) is the finite constant defined in \eqref{46a}.
\end{proposition}
\begin{IEEEproof}
Assume that there exists a stationary randomized policy \(\mathbf{w}^{\mathrm{st}}[n]\) that achieves the optimal time-average beamforming gain as
\begin{align}
\mathbb{E}\bigl[ \|\mathbf{H}[n]\mathbf{w}^{\mathrm{st}}[n]\|^2 \bigr] = \mathcal{G}_{\mathrm{opt}}, \label{eq:stationary_gain}
\end{align}
which satisfies the temperature constraints with a slack \(\epsilon > 0\)
\begin{align}
\mathbb{E}\bigl[ T_m[n] \mid \mathbf{Q}[n] \bigr] \leq T_{\text{th}} - \epsilon, \qquad \forall m. \label{eq:stationary_constraint}
\end{align}
The existence of such a policy is standard under the boundedness and feasibility assumption established in Lemma~\ref{lem:constraint_satisfaction}.
Based on the drift-plus-penalty bound in \eqref{46a} and the properties of the policy \(\mathbf{w}^{\mathrm{st}}[n]\), we have 
\begin{align}
\Delta(\mathbf{Q}[n]) &- V\,\mathbb{E}\bigl[ \|\mathbf{H}[n]\mathbf{w}[n]\|^2 \mid \mathbf{Q}[n] \bigr] \notag\\&\leq B - V \mathcal{G}_{\mathrm{opt}} - \epsilon \sum_{m=1}^{M} Q_m[n].
\end{align}
Taking expectations and summing over \(n = 1, \dots, N\) yields 
\begin{align}
\mathbb{E}[L(\mathbf{Q}&[N+1])] - \mathbb{E}[L(\mathbf{Q}[1])] - V\sum_{n=1}^{N} \mathbb{E}[\|\mathbf{H}[n]\mathbf{w}[n]\|^2] \notag \\
&\leq NB - NV\mathcal{G}_{\mathrm{opt}}- \epsilon \sum_{n=1}^{N} \sum_{m=1}^M \mathbb{E}[Q_m[n]].
\end{align}
Rearranging terms and using \(L(\mathbf{Q}[1]) = 0\) gives
\begin{align}
\frac{1}{N}\sum_{n=1}^{N} \mathbb{E}[\|\mathbf{H}[n]\mathbf{w}[n]\|^2] \geq & \mathcal{G}_{\mathrm{opt}}  - \frac{B}{V}+ \frac{ \mathbb{E}[L(\mathbf{Q}[N+1])]}{VN} \notag \\ & 
- \frac{\epsilon}{VN} \sum_{n=1}^{N} \sum_{m=1}^M \mathbb{E}[Q_m[n]].
\end{align}
Taking the limit as $N \to \infty$ and noting that both the queue-length term and the Lyapunov term on the right-hand side are non‑negative, we have
\begin{align}
\lim_{N \to \infty} \frac{1}{N}\sum_{n=1}^{N} \mathbb{E}\bigl[ \|\mathbf{H}[n]\mathbf{w}[n]\|^2 \bigr] \geq \mathcal{G}_{\mathrm{opt}} - \frac{B}{V}.
\end{align}
This completes the proof.
\end{IEEEproof}

Proposition \ref{Overall Performance Bound} implies that the performance of the proposed algorithm is governed by the control parameter \(V\). Specifically, the achievable received SNR grows nonlinearly as \(\mathcal{O}(1/V)\). When \(V\) is sufficiently large, the algorithm asymptotically approaches the optimal performance of the original problem $(\mathcal{P}1)$. This convergence, however, comes at the cost of an increased risk of violating the long-term temperature constraint. Thus, by tuning \(V\), a trade-off between EM exposure compliance and system performance can be effectively realized.

\section{Numerical Results And Discussion}\label{Section5}
This section presents numerical results to validate the EM and thermal models proposed and to evaluate the performance of our adaptive exposure-aware beamforming scheme. It is assumed that the UE is equipped with $N_t=4$ antennas and the BS is equipped with $N_r=64$ antennas. The carrier frequency is set as 30 GHz (wavelength \( \lambda = 0.01 \, \text{m} \)). The antenna elements at both the UE and the BS are spaced \(\lambda/2\) apart. The polarization vector of the receive antenna is along the $z$-axis. The BS is positioned at the origin with a height \(h_r =  5 \, \text{m} \). For EM exposure assessment, we considered a spherical head model with a diameter of ten wavelengths. The tissue parameters, including density and conductivity, follow the values listed in Table \ref{tab:params}. 
$M = 15$ sampling points are uniformly distributed along a horizontal circle on the surface of the model to obtain the exposure constraint  \eqref{temcon}. 
The center of the head model is located at ${\mathbf{p}}_u=\left[100 \, \text{m}, 100 \, \text{m}, 1.5 \, \text{m}\right]$, and the UE is initially placed at $\mathbf{p}_{t,1}={\mathbf{p}}_u+[0,d_{\text{ref}},0]$, where $d_{\text{ref}}$ is the reference distance. Time-varying EM exposure is analyzed over \(N = 3600\) time slots, where each time slot has a duration of \(\Delta t = 0.1 \, \text{s}\). During all time slots, the UE position is uniformly distributed within an annular region centered at \({\mathbf{p}}_u\), with inner radius \(d_{\min}\) and outer radius \(d_{\max}\). UE's tilt angle $\alpha_t$ and polar angle $\beta_t$ are randomly distributed over the ranges of $[-90^\circ, 90^\circ]$ and $[-20^\circ, 20^\circ]$, respectively. To address potential extreme exposure scenarios, a high transmit power constraint is adopted to explore the algorithmic performance boundaries \cite{pc}. Unless stated otherwise, maximum transmit power \( P_{\max} = 5 \, \text{W} \), noise variance \( \sigma^2 = 0.1 \), the antenna factor \(A_{F}=1\), the power transmission coefficient $T_{\mathrm{tr}}=0.8$ and the control parameter \( V = 5\times10^{-4} \).
\begin{figure}[tbp]
\centering
\includegraphics[width=0.48\textwidth]{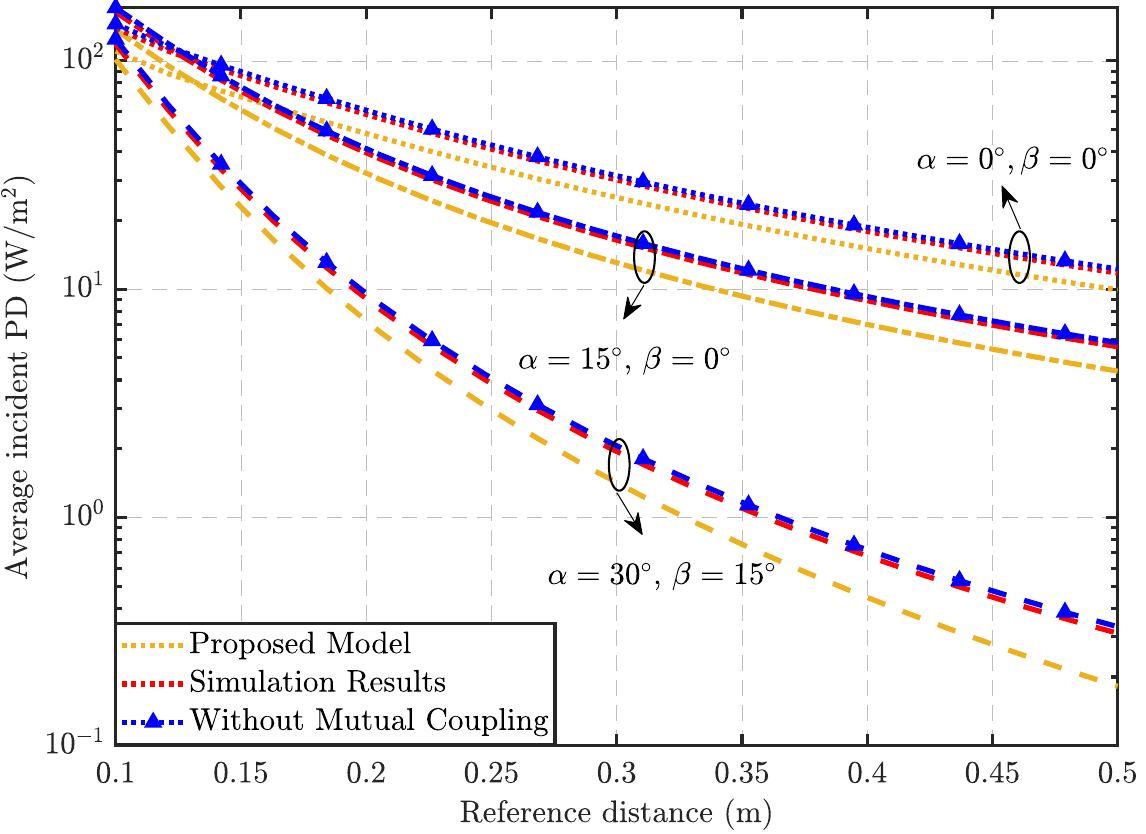}
\caption{Averaged power density versus the reference distance  $d_{\text{ref}}$ for different UE tilt and polar angles. }
\label{f3}
\end{figure}

To evaluate the performance of our proposed approach, we compare it with the following four baseline schemes:
\begin{itemize}
\item[$\bullet$] \textbf{Worst-case Power Back-off}: This conservative approach selects beamforming vectors based solely on the transmit power constraint.   The transmit power is reduced to a fixed level that satisfies the PD constraint under all possible UE orientations and positions, ensuring compliance in the worst-case scenario.

\item[$\bullet$] \textbf{Adaptive Power Back-off}: This method first determines the optimal beamformer using the available CSI, then dynamically adjusts the transmit power in each time slot based on real-time UE position and orientation knowledge to satisfy the instantaneous PD constraints.

\item[$\bullet$] \textbf{Per-slot Optimal}: This scheme uses numerical optimization via the fmincon toolbox to optimize the beamforming vector, maximizing the received SNR while under power and PD constraints in each time slot.  This approach yields the optimal solution for each individual slot without accounting for long‑term exposure accumulation.
\item[$\bullet$] \textbf{Unconstrained Optimal}:  This idealistic scheme selects beamforming vectors to maximize the received SNR without any EM exposure constraints, serving as a theoretical performance upper bound for comparison.
\end{itemize}
\begin{figure}[tbp]
\centering
\subfigure[Temporal evolution under $ V = 5\times10^{-4}$] 
{
\includegraphics[width=0.48\textwidth]{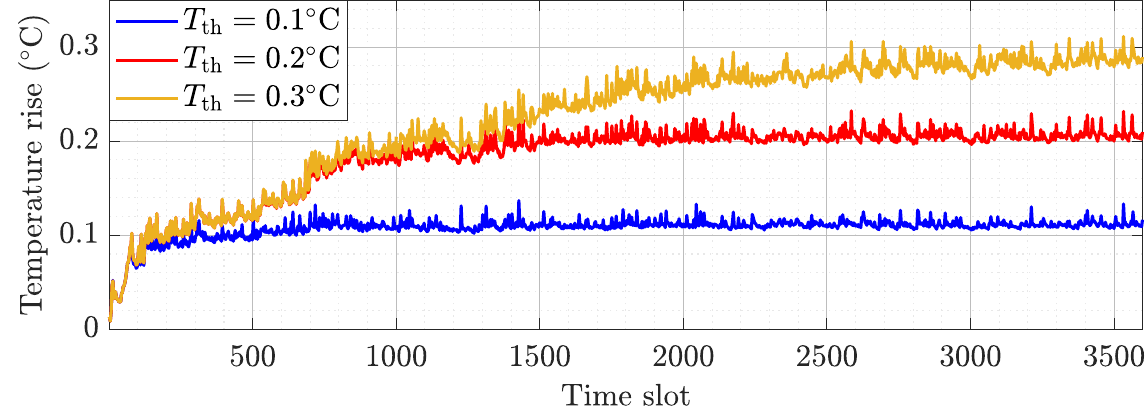}
\label{subfig:1}
}
\subfigure[Temporal evolution under $ V = 1\times10^{-4}$]{
\includegraphics[width=0.48\textwidth]{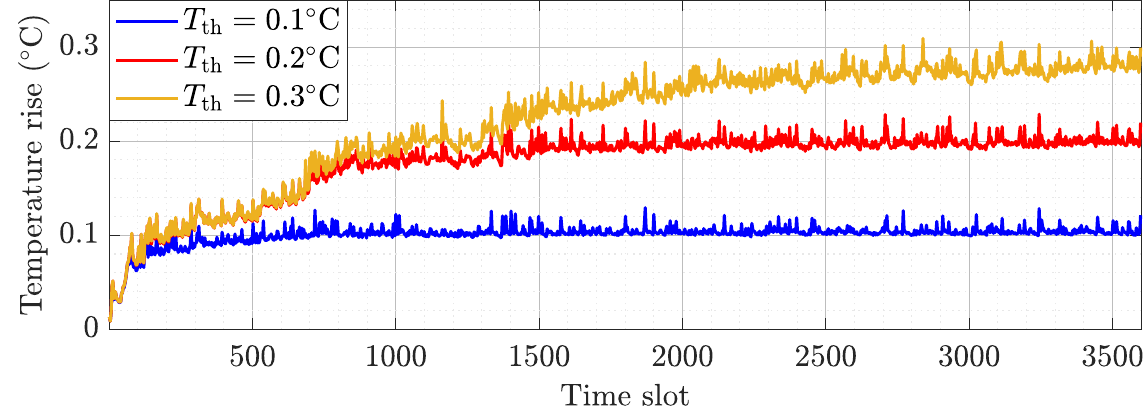}
\label{subfig:2}
}
\caption{Temporal evolution of average temperature under different control parameter \(V\) and temperature thresholds. 
}
\label{f4}
\vspace{-0.5 em}
\end{figure}

Fig. \ref{f3} shows the average incident PD over all sampling points versus the reference distance \(d_{\text{ref}}\) for different UE array orientations at the first time slot. The beamforming vector is initialized with equal weights. The distance \(d_{\text{ref}}\) varies from \(0.1\)$\text{ m}$ to \(0.5\)$\text{ m}$. 
The simulation results are obtained using an antenna toolbox, and the proposed model is calculated according to \eqref{eq:PD_calculation}. 
For comparison, a simplified reference model that ignores mutual coupling is also plotted, which only considers the impedance matrix containing the self-impedances of the isolated antenna elements
It can be observed that the proposed model matches the simulated PD accurately, confirming its validity for use in our subsequent beamforming design. 
In contrast, neglecting mutual coupling leads to an incorrect impedance matrix and consequently introduces error in the PD estimate. Furthermore, the UE orientation is shown to have a substantial impact on the exposure level. Thus, a beamforming strategy designed solely for a worst-case orientation would unnecessarily compromise the communication performance.

Fig. \ref{f4} shows the temporal evolution of the average temperature at sampling points under different control parameters \( V=5\times10^{-4} \) and \( 1\times10^{-4} \). To account for varying exposure constraints imposed by different scenarios and subjects, we set distinct temperature thresholds \(T_\text{max} = 0.1, 0.2\) and \(0.3^\circ\text{C}\).
In the early stage, since exposure conditions remain well below the thresholds, the system prioritizes seizing transmission opportunities over strict exposure compliance, thereby achieving near-optimal reception performance.
 Over time, the increasing weight assigned to EM exposure in the algorithm leads the system to suppress temperature rise progressively. Concurrently, heat dissipation through blood perfusion becomes significant and eventually balances the external thermal input, resulting in the average temperature converging to a specific critical threshold.

\begin{figure}[tbp]
\centering
\includegraphics[width=0.48\textwidth]{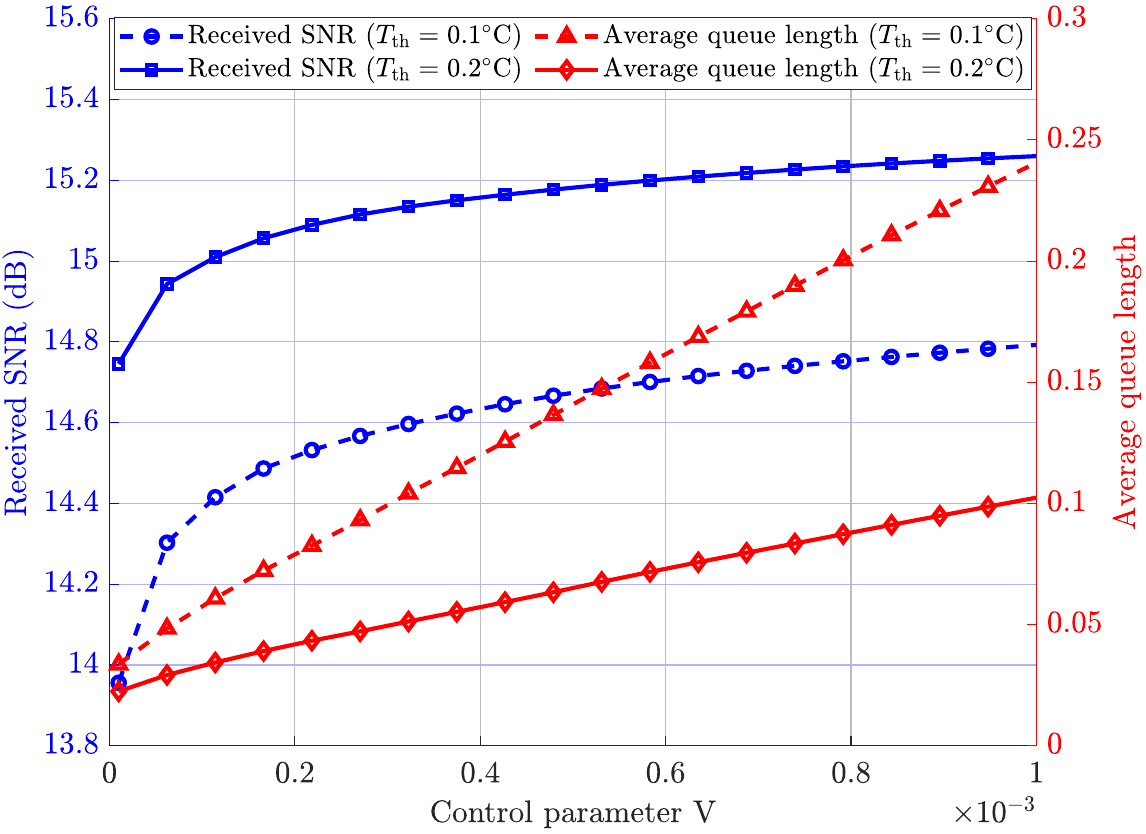}
\caption{Trade-off between the received SNR and average queue length versus control parameter for different temperature thresholds.}
\label{f5}
\end{figure}

In Fig. \ref{f5}, we depict the received SNR and the average virtual queue length versus the control parameter \( V \). 
The average queue length is proportional to the severity of the temperature exposure. When \( V \) is small, the algorithm prioritizes EM exposure safety, leading to stricter temperature control and hence a larger average queue length. As \( V \) increases, the algorithm progressively sacrifices EM exposure to maximize the SNR, until the performance eventually saturates.
Under the same control parameter \( V \), a lower temperature threshold, i.e., \( T_{\text{th}} = 0.1^\circ\text{C} \), implies a tighter exposure constraint and results in lower system performance. This is because the algorithm must more actively manage the virtual queues to ensure the stricter safety limit is met, which restricts the beamforming freedom and thus reduces the achievable SNR.
These results confirm that the control parameter \( V \) effectively governs the fundamental trade-off between exposure safety and communication performance in the proposed beamforming framework.

\begin{figure}[tbp]
\centering
\includegraphics[width=0.48\textwidth]{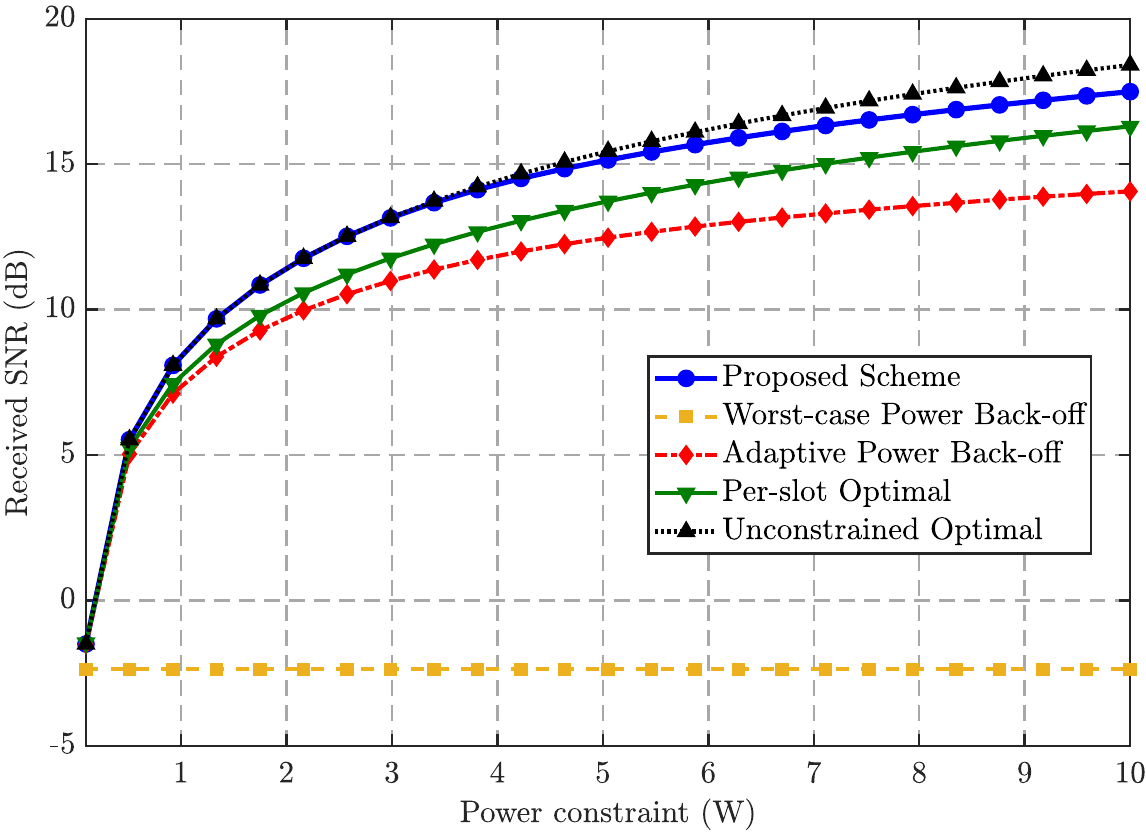}
\caption{Received SNR versus transmit power constraint for the proposed scheme and baseline schemes.}
\label{f6}
\end{figure}

Figure \ref{f6} compares the received SNR performance of the proposed scheme against four baseline schemes under different transmit power constraints. The PD constraint of four baseline schemes is given by $\mathcal{I}_{\text{th}}=20\text{W}/\text{m}^2$ \cite{r8}. As observed, our approach closely tracks the performance of the unconstrained optimal strategy, maintaining a performance gap of less than $10\%$ while satisfying EM exposure constraints.
The worst-case back-off scheme substantially degrades system performance since the exposure level varies significantly with changes in the device's position and orientation.
As power constraints increase, the Adaptive Power Back-off scheme requires significant power reduction to comply with instantaneous exposure limits.
Additionally, the proposed scheme achieves a 1–2 dB gain over the Per-slot Optimal scheme. This is because the proposed scheme overcomes the conservatism of instantaneous hard constraints, enabling time-domain adaptive resource optimization while satisfying the temperature constraint.

\begin{figure}[tbp]
\centering
\includegraphics[width=0.48\textwidth]{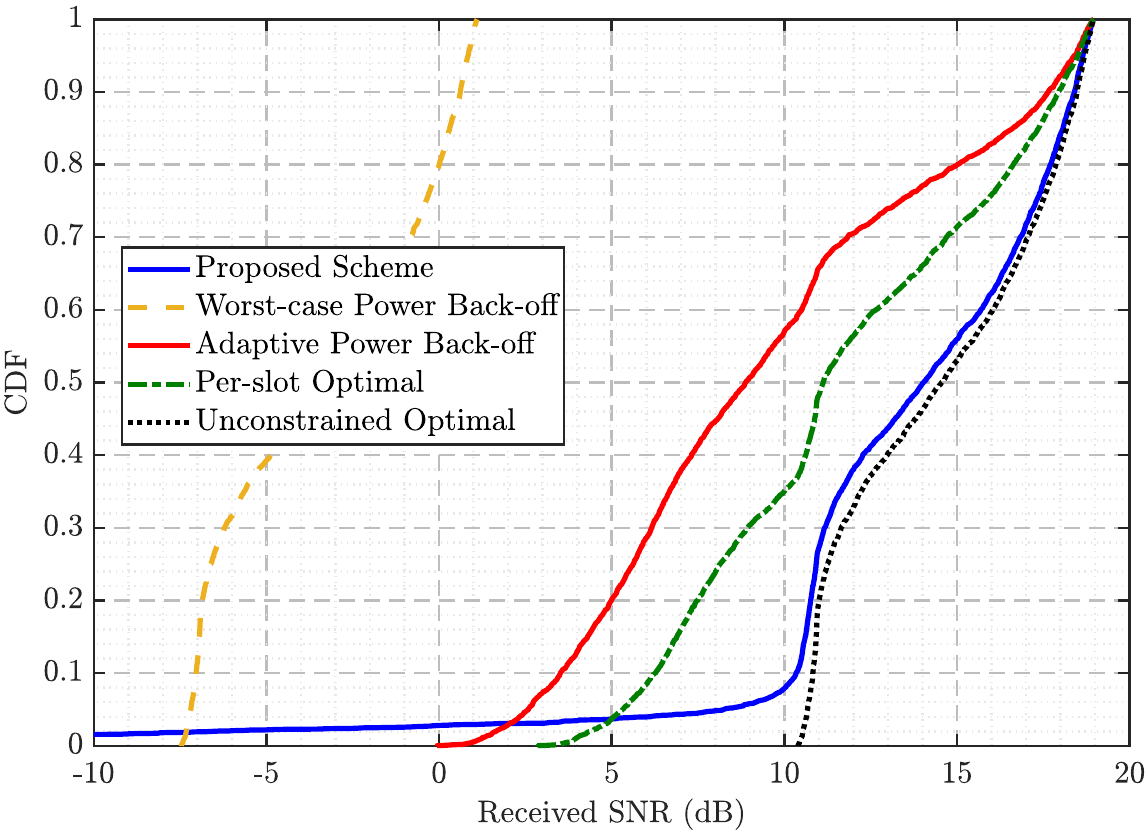}
\caption{Cumulative distribution of received SNR for different schemes.}
\label{f7}
\end{figure}

Fig. \ref{f7} presents the cumulative distribution function (CDF) of the received SNR over 3600 time slots for the considered schemes. The proposed scheme achieves superior SNR performance across most probability levels.
At the  $50\%$ probability level, our scheme attains an SNR of 14.1 dB, outperforming the per‑slot optimal scheme (11.1 dB). This improvement arises since the per‑slot optimal strategy adheres strictly to instantaneous safety constraints in every slot, even when the resultant temperature rise would not violate the long‑term threshold, thereby forgoing potential performance improvement.
A  Significant performance degradation is observed for the proposed scheme below the 2\% probability level. This is because the algorithm enters a silent mode when the exposure risk is high, which guarantees long‑term exposure safety.
Furthermore, the per‑slot optimal method requires solving a global optimization problem in each time slot, leading to a computationally prohibitive complexity for its practical deployment.

\begin{figure}[tbp]
\centering
\includegraphics[width=0.48\textwidth]{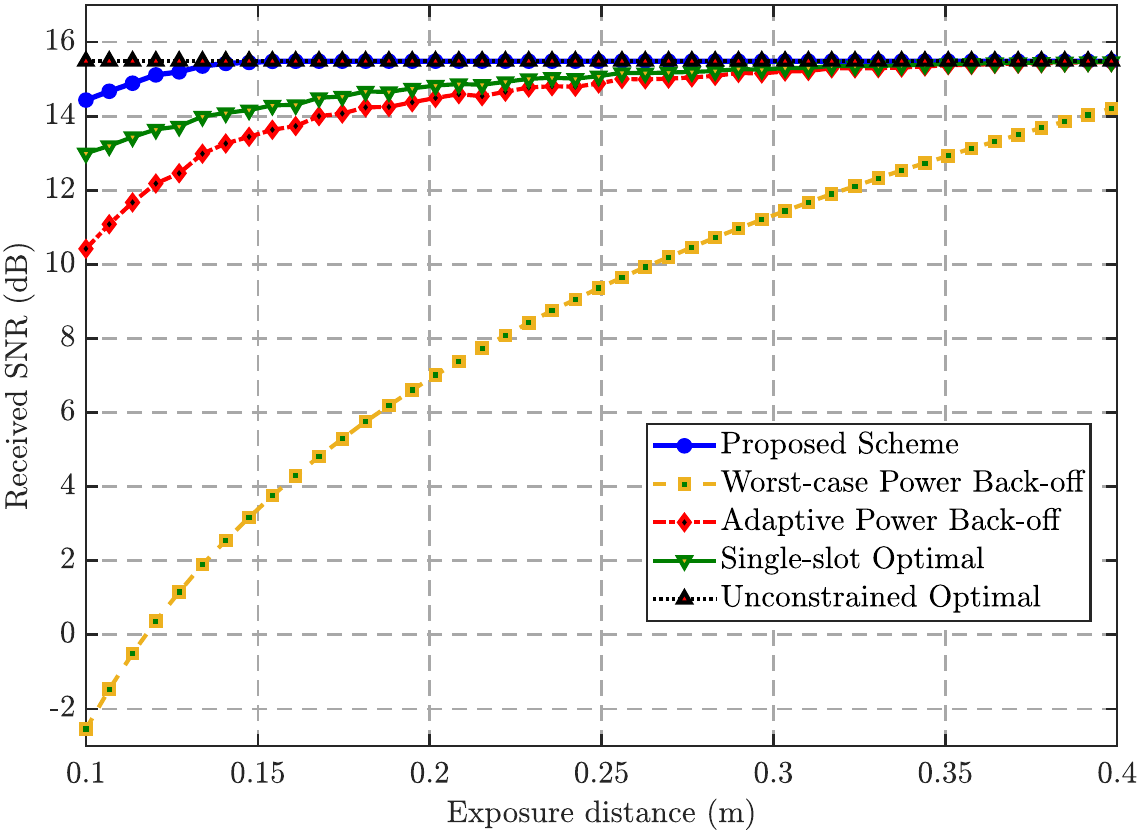}
\caption{Received SNR versus exposure distance \(d_{\text{exp}}\) for different schemes under different EM exposure conditions.}
\label{f8}
\end{figure}

To examine the performance under different EM exposure conditions, we assume the UE position is fixed on a circle centered at \(\mathbf{p}_u\) with a radius \(d_{\text{exp}}\), which we term the exposure distance and use to quantify the exposure intensity level. Figure \ref{f8} illustrates the received SNR as a function of \(d_{\text{exp}}\).
The worst-case power back-off scheme maintains a low SNR across all distance ranges. 
In contrast, both the adaptive power back‑off and the per‑slot optimal scheme exhibit gradual SNR improvement with increasing \(d_{\text{exp}}\), eventually approaching the optimal performance.
Under high-exposure scenarios, the proposed scheme surpasses the per‑slot optimal scheme and shows a clear advantage over other baselines. As the distance increases and exposure weakens, the performance of the proposed scheme improves rapidly and converges to that of the unconstrained optimal strategy. This asymptotic behavior indicates that the dominant system constraint shifts from the EM exposure limit to the transmit power budget.
Overall, the results demonstrate the robustness of the proposed scheme across a wide range of exposure intensities.

\section{Conclusion}\label{Section6}

In this paper, we have developed a long-term thermal EM exposure constraint model and proposed a novel adaptive exposure-aware beamforming design for a mmWave uplink system. A physically consistent EM radiation model has been developed to accurately characterize the channel and exposure metric, incorporating combined effects such as polarization, spherical wavefront, and mutual coupling.    
Based on this model and the BHTE, we have derived a closed-form thermal impulse response and characterized the temperature evolution in tissue under time-varying EM exposure. This analysis has revealed that exposure safety is governed by a long-term balance between EM power absorption and blood perfusion-driven cooling. Building on this insight, we have formulated an adaptive exposure-aware beamforming optimization problem that replaces rigid instantaneous constraints with a flexible average temperature constraint.     Using Lyapunov optimization theory, we have derived an optimal scheme with a closed-form solution requiring only instantaneous channel and exposure state information.   
Simulations have verified that the proposed algorithm stabilizes temperature and delivers near-optimal SNR. In contrast to conventional baselines, it dynamically manages the exposure budget and opportunistically boosts transmission when thermally permissible, thereby enhancing system performance.
\vspace{-0.05cm}
\section*{Appendix A\\Proof of Lemma \ref{aq1}}\label{app:radiation_field}\vspace{0.05cm}
The EM field radiated by a current distribution \(\mathbf{J}(\mathbf{r}')\) is derived from the magnetic vector potential  \(\mathbf{A}(\mathbf{r})\), which satisfies the inhomogeneous Helmholtz equation \cite{rm}:
\begin{align}
(\nabla^2 + k^2) \mathbf{A}(\mathbf{r}) = -\mu_0 \mathbf{J}(\mathbf{r'}),
\label{A1}
\end{align}
where \(\mu_0\) is the permeability of free space.  

Applying the distribution of current density in \eqref{eq6}. The solution to \eqref{A1} is given by the convolution of the current distribution with the scalar Green's function as
\begin{align}
\mathbf{A}(\mathbf{r}) = \mathbf{n}_t \frac{\mu_0}{4\pi} \int_{-h}^{h} I(l) \frac{e^{-jkR(l)}}{R(l)}  dl, 
\label{A4}
\end{align}
where $R(l) = \|\mathbf{r} - (\mathbf{r}_0 + l\mathbf{n}_t)\|$. Equation \eqref{A4} indicates that $\mathbf{A}(\mathbf{r})$ is parallel to the dipole axis, i.e., \(\mathbf{A}(\mathbf{r}) = A_{\parallel}(\mathbf{r}) \mathbf{n}_t\), where \(A_{\parallel} = \mathbf{n}_t \cdot \mathbf{A}\).

Under the Lorenz gauge, the electric field components  parallel and perpendicular to the dipole axis \(\mathbf{n}_t\) are derived
from the vector potential as
\begin{align}
j\omega \mu_0 \epsilon_0 \mathcal{E}_{\parallel} &= \partial_{\parallel}^2 A_{\parallel} + k^2 A_{\parallel}, \label{A5} \\
j\omega \mu_0 \epsilon_0 \mathcal{E}_{\perp} &= \partial_{\perp} \partial_{\parallel} A_{\parallel}, \label{A6}
\end{align}
where \(\partial_{\parallel} = \mathbf{n}_t \cdot \nabla\) and \(\partial_{\perp}\) denotes the spatial derivative perpendicular to \(\mathbf{n}_t\). Using the free-space impedance \(\eta = \sqrt{\mu_0/\epsilon_0}\) and the relation \(\omega \mu_0 \epsilon_0 = k / \eta\), \eqref{A5} and \eqref{A6} simplify to 
\begin{align}
\mathcal{E}_{\parallel} &= \frac{j\eta}{k} \left( \partial_{\parallel}^2 A_{\parallel} + k^2 A_{\parallel} \right), \label{A7} \\
\mathcal{E}_{\perp} &= \frac{j\eta}{k} \partial_{\perp} \partial_{\parallel} A_{\parallel}. \label{A8}
\end{align}

Substituting \(A_{\parallel}\) from \eqref{A4} into \eqref{A7} yields
\begin{align}
\partial_{\parallel}^2 A_{\parallel} + k^2 A_{\parallel} = \frac{\mu_0}{4\pi} \int_{-h}^{h} I(l) \left( \partial_{l}^2 + k^2 \right) G(R(l))  dl,
\label{A9}
\end{align}
where we denote the Green’s function \(G = e^{-jkR(l)}/R(l)\) and replace $\partial_{\parallel}^2$ by $\partial_{l}^2$. Since \(I(l)\) satisfies the one-dimensional Helmholtz equation, i.e. \(\left(\partial_{l}^2 + k^2\right) I(l) = 0\), the integrand in \eqref{A9} can be expressed as a total derivative using the identity
\begin{align}
I(l)\left(\partial_{l}^2 + k^2\right)G = \partial_{l} \left[ I(l) \partial_{l} G - G \partial_{l} I(l) \right].
\label{A10}
\end{align}
Substituting \eqref{A10} into \eqref{A9} yields
\begin{align}
\partial_{\parallel}^2 A_{\parallel} + k^2 A_{\parallel} = \frac{\mu_0}{4\pi} \left( \int_{-h}^{h} \partial_{l}\left[I \partial_{l} G\right] dl - \int_{-h}^{h} \partial_{l}\left[G \partial_{l} I\right] dl \right).
\label{A11}
\end{align}

The first integral vanishes since the end condition \(I(\pm h) = 0\). Evaluating over \([-h,0]\) and \([0, h]\) with \(\partial_{l} I(0^{\pm}) = \mp k I_m \cos(kh)\) and \(\partial_{l} I(\pm h) = \mp k I_m\), The second integral is rewritten as
\begin{align}
\!\int_{-h}^{h} \!\partial_{l}\left[G \partial_{l} I\right] dl \!=\! k I_m \left[ 2\cos(kh) G(R_0) \!-\! G(R_1) \!-\! G(R_2) \right],
\label{A12}
\end{align}
where
\[
R_0 \!=\! \|\mathbf{r}-\mathbf{r}_0\|,\quad
R_1 \!=\! \|\mathbf{r}-(\mathbf{r}_0 + h\mathbf{n}_t)\|,\quad
R_2 \!=\! \|\mathbf{r}-(\mathbf{r}_0 - h\mathbf{n}_t)\|.
\]
Combining \eqref{A11}, \eqref{A12}, and \eqref{A7}, we give
\begin{align}
\mathcal{E}_{\parallel}(\mathbf{r}) \!=\! -\frac{j\eta I_m}{4\pi} \left( \frac{e^{-jkR_1}}{R_1} \!+\! \frac{e^{-jkR_2}}{R_2} \!-\! 2\cos(kh) \frac{e^{-jkR_0}}{R_0} \right).
\label{A13}
\end{align}

The transverse component \(\mathcal{E}_{\perp}\) can be efficiently derived via the magnetic field. In a local cylindrical coordinate system \((\rho, \phi, z)\) aligned with the dipole (where \(z\) is along \(\mathbf{n}_t\)), Ampère's circuital law gives:
\begin{align}
\frac{1}{\rho} \frac{\partial}{\partial \rho} (\rho H_\phi) = j\omega \epsilon_0 \mathcal{E}_{\parallel}.
\label{eq:ampere_circuital}
\end{align}
Substituting \(\mathcal{E}_{\parallel}\) from \eqref{A13} into \eqref{eq:ampere_circuital} and integrating, we find the magnetic field:
\begin{align}
H_\phi(\mathbf{r}) = \frac{j I_m}{4\pi \rho} \left( e^{-jkR_1} + e^{-jkR_2} - 2\cos(kh) e^{-jkR_0} \right),
\label{eq:H_phi_exact}
\end{align}
where \(\rho = \| (\mathbf{r} - \mathbf{r}_0) - [(\mathbf{r} - \mathbf{r}_0) \cdot \mathbf{n}_t] \mathbf{n}_t \|\) is the perpendicular distance from the observation point to the dipole axis.

 When \(d = \|\mathbf{r} - \mathbf{r}_0\| \gg \lambda\), we apply the standard approximations: For amplitude terms, we set \(R_1 \approx R_2 \approx R_0 \approx d\). For phase terms, we retain first-order expansions to preserve wavefront coherence:
\begin{align}
R_{{1}} &= \|\mathbf{r} - \mathbf{r}_0 - h \mathbf{n}_t\| \approx d - h (\hat{\mathbf{r}} \cdot \mathbf{n}_t), \\
R_2 &= \|\mathbf{r} - \mathbf{r}_0 + h \mathbf{n}_t\| \approx d + h (\hat{\mathbf{r}} \cdot \mathbf{n}_t),
\end{align}
where \(\hat{\mathbf{r}} = (\mathbf{r} - \mathbf{r}_0)/d\). Defining the angle \(\psi\) between \(\mathbf{n}_t\) and \(\hat{\mathbf{r}}\) via \(\cos \psi = \mathbf{n}_t \cdot \hat{\mathbf{r}}\), we have \(\rho = d \sin \psi\). Applying these approximations to \eqref{A13} and \eqref{eq:H_phi_exact} yields the expressions:
\begin{align}
\mathcal{E}_{\parallel}(\mathbf{r}) &\approx -\frac{j\eta I_m}{2\pi d} e^{-jkd} \left[ \cos(kh \cos\psi) - \cos(kh) \right], \label{eq:E_parallel_far} \\
H_\phi(\mathbf{r}) &\approx \frac{j I_m e^{-jkd}}{2\pi d \sin\psi} \left[ \cos(kh \cos\psi) - \cos(kh) \right]. \label{eq:H_phi_far}
\end{align}

The transverse electric field follows from Faraday's law, \(j\omega \epsilon_0 \mathcal{E}_{\perp} = -\partial_{\parallel} H_\phi\). Using the approximation \(\partial_{\parallel}(e^{-jkd}) \approx -jk \cos\psi \, e^{-jkd}\), we obtain:
\begin{align}
\mathcal{E}_{\perp}(\mathbf{r}) &\approx \frac{j \eta I_m e^{-jkd}}{2\pi d} \frac{\cos\psi}{\sin\psi} \left[ \cos(kh \cos\psi) - \cos(kh) \right]. \label{eq:E_perp_far}
\end{align}

The total electric field vector is the superposition \(\boldsymbol{\mathcal{E}} = \mathcal{E}_{\parallel} \mathbf{n}_t + \mathcal{E}_{\perp} \hat{\boldsymbol{e}}\), where \(\hat{\boldsymbol{e}}\) is the unit vector in the direction of increasing \(\rho\) (perpendicular to \(\mathbf{n}_t\) and lying in the plane containing \(\mathbf{n}_t\) and \(\hat{\mathbf{r}}\)). Expressing \(\hat{\boldsymbol{e}}\) in terms of \(\hat{\mathbf{r}}\) and \(\mathbf{n}_t\) as \(\hat{\boldsymbol{e}} = (\hat{\mathbf{r}} - \cos\psi \, \mathbf{n}_t)/\sin\psi\) and combining \eqref{eq:E_parallel_far} and \eqref{eq:E_perp_far}, we arrive at the compact expression:
\begin{align}
\boldsymbol{\mathcal{E}}(\mathbf{r}) \approx \frac{j \eta I_m}{2\pi d} \frac{\cos(kh \cos\psi) - \cos(kh)}{\sin\psi} e^{-jkd} \, \hat{\boldsymbol{\rho}}(\hat{\mathbf{r}}),
\label{eq:total_E_far}
\end{align}
where \(\hat{\boldsymbol{\rho}}(\hat{\mathbf{r}}) = \left[ (\mathbf{n}_t \cdot \hat{\mathbf{r}}) \hat{\mathbf{r}} - \mathbf{n}_t \right] / \sin\psi\) is the unit polarization vector orthogonal to \(\hat{\mathbf{r}}\). This completes the derivation.

\vspace{-0.3cm}
\section*{Appendix B\\Proof of Lemma \ref{aq21}}
\label{app: Mutual Impedance}
When antenna-$q$ is driven by the current \(I_q\) while antenna-$p$ is open-circuited, the field radiated by antenna-$q$ induces an open-circuit voltage \(V_{pq,\mathrm{oc}}\) on antenna-$p$. The mutual impedance is defined as
\begin{align}
Z_{pq} = \frac{V_{pq,\mathrm{oc}}}{I_q},
\label{eq:mupz}
\end{align}

According to the reciprocity theorem, the open-circuit voltage induced on the antenna-$q$ can be expressed as
\begin{align}
V_{pq,\mathrm{oc}} = -\frac{1}{I_p} \int_{-h_p}^{h_p} \boldsymbol{\mathcal{E}}_{pq}(l) \cdot \mathbf{n}_t  I_p(l) dl,
\label{eq:Voc}
\end{align}
where $I_p=I_p(0)$ is the input current of antenna-$p$ when it is transmitting, $\boldsymbol{\mathcal{E}}_{pq}(l)$ is the electric field vector produced by antenna-$q$ at the observation point $\mathbf{r}_p+l \mathbf{n}_t$
on antenna-$p$ and the term $\boldsymbol{\mathcal{E}}_{pq}(l) \cdot \mathbf{n}_t$  gives the parallel component of the electric field along the antenna axis $\mathbf{n}_t$, i.e., $\mathcal{E}_{\parallel}=\boldsymbol{\mathcal{E}}_{pq}(l) \cdot \mathbf{n}_t$

Adopting the thin-wire approximation, the current distributions on antennas are assumed to be sinusoidal as
\begin{align}
I_q(l') = I_q \frac{\sin\bigl(k(h_q - |l'|)\bigr)}{\sin(k h_q)},
I_p(l) = I_p \frac{\sin\bigl(k(h_p - |l|)\bigr)}{\sin(k h_p)}, \label{eq:I_q}
\end{align}
where $l^{\prime} \in\left[-h_q, h_q\right]$ and $l \in\left[-h_p, h_p\right]$ are local coordinates along the antennas. Similar to the derivation in \eqref{A13} of Appendix A, the axial component of the electric field $\mathcal{E}_{\parallel}$ is 
\begin{align}
\mathcal{E}_{\parallel}&=\boldsymbol{\mathcal{E}}_{pq}(l) \cdot \mathbf{n}_t \label{eq:Efield} \\&= -\frac{j\eta I_{p}}{4\pi\sin (k h_q)}
\Biggl[ \frac{e^{-jkR_{{1}}}}{R_{{1}}} \!+\! \frac{e^{-jkR_2}}{R_2} \!-\! 2\cos(k h_q) \frac{e^{-jkR_0}}{R_0} \Biggr],\notag
\end{align}
where the distances $R_0,R_{{1}},R_2$ are defined as 
\begin{align}
R_0 &= \| \mathbf{r}_p - \mathbf{r}_q + l \mathbf{n}_t \|, \nonumber \\
R_{{1}} &= \| \mathbf{r}_p - \mathbf{r}_q + (l - h_q) \mathbf{n}_t \|, \\
R_2 &= \| \mathbf{r}_p - \mathbf{r}_q + (l + h_q) \mathbf{n}_t \|. \nonumber\label{eq:R_def}
\end{align}
These distances correspond, respectively, to the separation between the observation point on antenna-$p$ and the center, the top end, and the bottom end of antenna-$q$.

 Substituting  \eqref{eq:Efield} and  \eqref{eq:Voc} into \eqref{eq:mupz} yields
\begin{align}
Z_{pq} &\!= \!-\frac{1}{I_p I_q} \int_{-h_p}^{h_p} \boldsymbol{\mathcal{E}}_{pq}(l) \cdot \mathbf{n}_t \, I_p(l)  \, dl \notag \\
 &\!=\! \frac{j\eta}{4\pi \sin (k h_p) \sin (k h_q)} \!\int_{-h_p}^{h_p} F_{pq}(l)  \, dl,
\end{align}
with the kernel function
\begin{align}
\!F_{pq}(l) \!=\! \left[\frac{e^{-j k R_{1}}}{R_{1}}\!+\!\frac{e^{-j k R_{2}}}{R_{2}}\!-\!2\! \cos (k h_q) \frac{e^{-j k R_{0}}}{R_{0}}\right]\!\sin\bigl[k(h_p \!-\! |l|)\bigr].\notag
\end{align}

\bibliographystyle{IEEEtran}
\bibliography{main.bib}
\end{document}